\documentclass[12pt,twoside,openany]{book}

\usepackage[letterpaper,lmargin=2.9cm,rmargin=1.9cm,bmargin=2.8cm,tmargin=2.8cm,includeheadfoot]{geometry} 

\usepackage{graphicx}  
\usepackage{tikz, tikz-3dplot}
\usepackage{flafter}  
\usepackage[list=off,font=small,skip=3pt]{caption}
\usepackage{subcaption}

\usepackage{lipsum}

\usepackage{cite}

\usepackage{amsthm}
\usepackage{amsmath,amssymb}  
\usepackage{textcomp} 
\usepackage{bm}  

\usepackage[utf8]{inputenc} 
\usepackage[T1]{fontenc}
\usepackage{lmodern}
\usepackage{fix-cm}
\usepackage[spanish,slovene]{babel}

\usepackage{booktabs}
\renewcommand{\arraystretch}{1} 

\usepackage{enumitem}
\setlist{itemsep=0.5pt}

\usepackage[plain]{fancyref}
\usepackage{hyperref}
\hypersetup{
    colorlinks=true,
    linkcolor=black,
    citecolor=black,
    filecolor=black,
    urlcolor=black
}

\usepackage{fancyhdr}
\usepackage{aas_macros} 
\usepackage[]{tocbibind} 

\usepackage{comment}

\makeatletter 
\DeclareRobustCommand*{\bfseries}{%
  \not@math@alphabet\bfseries\mathbf
  \fontseries\bfdefault\selectfont
  \boldmath
}
\makeatother

\begin{document}
	
	\selectlanguage{spanish}

    \renewcommand{\tablename}{Tabla}

    \renewcommand{\thefootnote}{\Roman{footnote}}

    \lefthyphenmin=2
    \righthyphenmin=2

    \frontmatter

\thispagestyle{empty}
\newlength{\titlerulewidth}
\setlength{\titlerulewidth}{0.4pt} 

\begin{center}
	
	\vspace*{-0.1cm} 
	
	{Instituto Superior de Tecnolog\'ias y Ciencias Aplicadas, Universidad de La Habana.}
	
	{Departamento de F\'isica Nuclear.}
	
	{Instituto de Cibern\'etica, Matem\'atica y Física.}
	
	\vspace{0.2cm} 

    \includegraphics[width=2.9in,keepaspectratio]{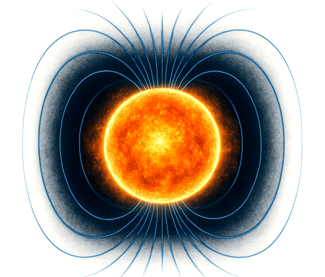}
	
	\vspace{0.2cm} 
	
	{\large TESIS}
	
	\vspace{0.15cm} 
	
	{\large presentada en opci\'on al grado científico de}
	
	\vspace{0.15cm} 
	
	{\large Licenciado en Física Nuclear}
	
	\vspace{0.25cm} 
	
	\noindent\rule{\textwidth}{\titlerulewidth} 
	
	\vspace{0.25cm} 
	{\fontsize{16}{18}\selectfont\bfseries 
		\begin{tabular}{@{}c@{}}
			EFECTO DEL CAMPO MAGN\'eTICO EN ESTRELLAS\\[0.2cm] 
			DE BOSONES ESCALARES CARGADOS
		\end{tabular}
	}
	\vspace{0.25cm} 
	
	\noindent\rule{\textwidth}{\titlerulewidth} 
	
	\vspace{1.0cm} 
	
	\begin{tabular}{rl}
		\textbf{Autor:} & Marcos Alejandro Alvarez Hern\'andez \\
		\textbf{Tutora:} & Dra. Aurora P\'erez Martínez \\
	\end{tabular}
	
	\vspace{1.0cm}
	
	\includegraphics[width=1.9in,keepaspectratio]{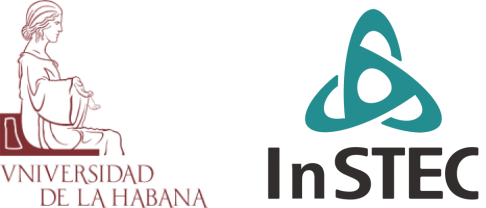}
	\hspace{0.5cm}
	\includegraphics[width=0.75in,keepaspectratio]{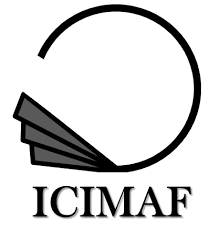}
	
	\vspace{0.15cm}
	
	{\large La Habana, 2025}
	
\end{center}

	\begin{center}
	\section*{Agradecimientos}
\end{center}


\begin{quote}
Al llegar al final de esta etapa tan significativa en mi vida, quiero expresar mi m\'as profundo agradecimiento a todas las personas que han sido parte esencial de este camino.
\begin{flushright}
    A mis padres, por su amor incondicional, por enseñarme con su ejemplo el valor del esfuerzo, la honestidad y la perseverancia. Gracias por creer en mí incluso en los momentos en que yo mismo dud\'e. 
    
    Al resto de mi enorme familia, por su apoyo constante y por estar siempre cerca, brind\'andome \'animo, comprensi\'on y un refugio al que siempre puedo volver.

    A los amigos que estuvieron a mi lado en cada paso del camino, por su compañía sincera, por hacer m\'as llevaderos los días difíciles y por recordarme siempre que no estoy solo.

    A mis profesores, quienes han sido una parte esencial de mi formaci\'on, no solo acad\'emica, sino tambi\'en personal. Gracias por transmitirme conocimientos, pero tambi\'en valores, perspectivas y formas de enfrentar la complejidad con rigor, sentido cr\'itico y curiosidad. 
    
    A los miembros del Departamento de Física Te\'orica del ICIMAF, por haberme acogido y ofrecerme un espacio de formaci\'on y crecimiento donde he aprendido no solo ciencia, sino tambi\'en colaboraci\'on, compromiso y comunidad.
    
    A mi tutora, gu\'ia acad\'emica inmejorable, por su pacientcia, por su entusiasmo contagioso y por creer en mí. Gracias por estar presente en cada etapa del proceso y motivarme a dar lo mejor de mí.
    
    A todos ustedes: gracias por haberme acompañado hasta aquí. Este logro tambi\'en les pertenece.

\end{flushright}

\end{quote}

\vfill

\pagestyle{plain}

    \pagestyle{empty}

\vfill

\begin{center}
\section*{Resumen}
\addcontentsline{toc}{chapter}{Resumen}
\end{center}
\medskip

\begin{quote}
	
En esta tesis se estudia la estructura y las propiedades termodin\'amicas de estrellas de Condensado de Bose-Einstein (EBE) en presencia de campos magn\'eticos intensos. Estos objetos, han surgido como alternativas te\'oricas plausibles a las estrellas de neutrones convencionales, especialmente en el contexto de observaciones astrof\'isicas que sugieren la existencia de fases no bari\'onicas en los n\'ucleos estelares. El trabajo aborda el estudio termodin\'amico de gases de bosones escalares cargados y fermiones (electrones, protones y neutrones) en presencia de un campo magn\'etico uniforme, prestando particular atenci\'on al l\'imite de bajas temperaturas. Se derivan expresiones para el potencial termodin\'amico, densidades de part\'iculas, energ\'ias, presiones anisotr\'opicas y magnetizaci\'on, incluyendo tanto contribuciones estad\'isticas como del vac\'io y separando esta \'ultima en la contribuci\'on del nivel de Landau m\'as bajo (LLL) y de los niveles excitados. A partir de estas expresiones, se construyen ecuaciones de estado (EdE) anisotr\'opicas que incorporan condiciones de equilibrio qu\'imico y de neutralidad de carga para un gas mixto \textit{npe}+$\pi^-$. Estas EdE se integran en un sistema de ecuaciones de estructura anisotr\'opicas basado en una m\'etrica axisim\'etrica, lo cual permite analizar c\'omo el campo magn\'etico modifica la estructura de las EBE, incluyendo sus masas, radios, deformaciones, corrimientos al rojo gravitacional y momento cuadrupolar de masa. Se presentan resultados num\'ericos que muestran la influencia del campo magn\'etico en las propiedades globales de estas estrellas, incluyendo posibles firmas observables. Este estudio representa, hasta donde se conoce, el primer an\'alisis detallado de EBE compuestas por bosones cargados en condiciones astrof\'isicamente realistas, contribuyendo al desarrollo de modelos te\'oricos consistentes para el interior de objetos compactos magnetizados.

\end{quote}

\vfill

\begin{center}
\section*{Abstract}
\addcontentsline{toc}{chapter}{Abstract}
\end{center}
\medskip

\begin{quote}

This thesis investigates the structure and thermodynamic properties of Bose-Einstein Condensate Stars (BEC stars) in the presence of intense magnetic fields. These objects have emerged as theoretically plausible alternatives to conventional neutron stars, especially in the context of astrophysical observations suggesting the existence of non-baryonic phases in stellar cores. The work focuses on the thermodynamic study of gases composed of charged scalar bosons and fermions (electrons, protons, and neutrons) under a uniform magnetic field, paying particular attention to the low-temperature limit. Expressions are derived for the thermodynamic potential, particle densities, energies, anisotropic pressures, and magnetization, including both statistical and vacuum contributions, with the latter separated into the contribution from the Lowest Landau Level (LLL) and excited levels. Based on these results, anisotropic equations of state (EoS) are constructed that incorporate chemical equilibrium and charge neutrality conditions for a mixed \textit{npe}+$\pi^-$ gas. These EoS are then implemented in a system of anisotropic structure equations based on an axisymmetric metric, allowing for the analysis of how the magnetic field modifies the structure of BEC stars, including their masses, radii, deformations, gravitational redshifts, and mass quadrupole moments. Numerical results are presented showing the influence of the magnetic field on the global properties of these stars, including potential observable signatures. This study represents, to the best of our knowledge, the first detailed analysis of BEC stars composed of charged bosons under astrophysically realistic conditions, contributing to the development of consistent theoretical models for the interior of magnetized compact objects.

\end{quote}

\vfill

    \tableofcontents


    \mainmatter

    \pagestyle{fancy}
    \pagestyle{plain}
    \chapter*{Introducci\'on} \label{intro}
\addcontentsline{toc}{chapter}{Introducci\'on}

Cuando una estrella agota su combustible nuclear, puede colapsar y dar lugar a un objeto compacto (OC)\cite{camenzind2007compact}. Con masas del orden de la del Sol, confinados en unas pocas decenas de kil\'ometros de radio, los OCs representan laboratorios naturales donde las leyes de la física se ponen a prueba bajo condiciones extremas de densidad y gravedad. Este grupo incluye enanas blancas (EB), estrellas de neutrones (EN) y agujeros negros (AN), cada uno formado a trav\'es de los estadíos finales evolutivos del colapso estelar y distinguido por los mecanismos de presi\'on de degeneraci\'on o fuerzas gravitacionales que evitan un mayor colapso. 


Entre los OCs, las ENs son particularmente interesantes al ser los \'unicos objetos conocidos en cuyo interior se puede alcanzar y superar la densidad de los n\'ucleos at\'omicos ($\rho_{nuc} = 2.7 \times 10^{14}$g$/$cm$^3$\cite{camenzind2007compact, NSObserations}). Si bien el nombre de Estrella de Neutrones tiene un origen hist\'orico, pues la primera explicaci\'on para la estabilidad de estos objetos sugería que estaban sostenidos por la presi\'on de un gas degenerado de fermiones masivos que, en ese momento, se propuso que fueran neutrones; hoy se sabe que los datos observacionales no pueden ser explicados por completo por esta composici\'on \'unicamente. 

Para calcular los perfiles y propiedades de las ENs, necesitamos informaci\'on sobre su estructura interna a trav\'es de ecuaciones de estado (EdE). Hasta ahora, carecemos de una EdE universalmente aceptada para las ENs. En su lugar, existen numerosos modelos que predicen diferentes propiedades. Los datos observados de las ENs pueden usarse para imponer restricciones a estos modelos.

Toda descripci\'on te\'orica de las ENs parte de un gas de neutrones con una peque\~na fracci\'on de protones y electrones, el gas \textit{npe} , de modo que se satisfagan la neutralidad de carga y el equilibrio beta\cite{camenzind2007compact,Schmitt2010}. A partir de este gas, diversos modelos para la composici\'on interna de la estrella se construyen a\~nadiendo interacciones o nuevas part\'iculas, dando lugar cada combinaci\'on a EdE distintas \cite{camenzind2007compact}. Otros modelos para el interior de las ENs se obtienen, en cambio, de suponer que sus condiciones internas son tales que favorecen la aparici\'on de nuevas fases de la materia. Tal es el caso de las Estrellas de Quarks o Estrellas Extra\~nas\cite{Terrero2021}, Estrellas de Bosones o de las Estrellas de condensado de Bose-Einstein\cite{Chavanis2012}. Igualmente, existen modelos llamados de  Estrellas H\'ibridas, en los que al n\'ucleo de la EN se le supone una estructura a capas. Por ejemplo, el n\'ucleo podr\'ia estar compuesto a su vez por un n\'ucleo de quarks deconfinados envuelto en una corteza de materia nuclear \cite{Masquerade}, pero podr\'ia tambien contener capas intermedias con mezcla de las dos fases \cite{Mourelle2024}.

Dentro del espectro de modelos, las estrellas de bosones, y en particular las estrellas de condensado de Bose-Einstein (EBE) han ganado relevancia tanto te\'orica como fenomenol\'ogica \cite{Liebling2023}. Estas configuraciones hipot\'eticas son localizados, carecen de horizonte de eventos y est\'an compuestos por materia bos\'onica autogravitante. Su estabilidad no depende de procesos termodin\'amicos ni de la degeneraci\'on de fermiones, sino de la presi\'on cu\'antica generada por el principio de incertidumbre de Heisenberg aplicado a un condensado de bosones en su estado fundamental \cite{Brito2024}. La configuraci\'on de la estrella depende del tipo de bosones, sus interacciones, masa, entre otros factores. 

Desde un punto de vista te\'orico, las EBEs son especialmente atractivas porque surgen en marcos físicos bien fundamentados y cuentan con un mecanismo de formaci\'on plausible: el enfriamiento gravitacional \cite{Liebling2023}. Observacionalmente, se han propuesto como candidatos para explicar aglomeraciones de materia oscura ultraligera \cite{Freitas2021}, e incluso como alternativas o complementos a los agujeros negros astrofísicos \cite{herdeiro2023}.

La mayoría de los objetos compactos conocidos presentan altos campos magn\'eticos, cuyo origen sigue sin poder explicarse. Estos campos magn\'eticos los alejan de la forma esf\'erica\cite{DaryelModelingCO} pues provoca que la presi\'on ejercida por el gas que compone la estrella sea distinta en las direcciones paralela y perpendicular al eje magn\'etico, dando lugar a ecuaciones de estado (EdE) anisotr\'opicas. El equilibrio macrosc\'opico de una estrella se logra gracias al balance entre la gravedad –que tiende a comprimir la materia que la compone hacia su centro– y la presi\'on que esta materia ejerce hacia afuera. Por tanto, un objeto compacto cuya presi\'on interna sea isotr\'opica es esf\'erico, mientras que en caso contrario la forma est\'a determinada por las diferencias entre las presiones.

En trabajos previos del grupo de investigaci\'on \cite{GQ21AnyT, QuinteroAngulo2022, GQ23FiniteT}, se investig\'o el comportamiento termodin\'amico de un gas de bosones vectoriales neutros magnetizado (GBVM) a temperaturas arbitrarias, y se deriv\'o la correspondiente ecuaci\'on de estado (EoS) con el objetivo de modelar EBE como una alternativa a las estrellas de neutrones convencionales.

En este trabajo nos proponemos complementar los resultados de estos trabajos, con el estudio de estrellas de bosones escalares cargados. Esto constituye, hasta donde sabemos, el primer estudio de la contraparte bos\'onica cargada en tales condiciones, De acuerdo con esto, y teniendo en cuenta que la construcci\'on de modelos te\'oricos consistentes es esencial para la interpretaci\'on de las observaciones astrof\'isicas, trabajaremos bajo la hip\'otesis de que un gas de bosones escalares cargados magnetizados (GBECM) en condiciones de baja temperatura y alta densidad puede formar un condensado de Bose-Einstein (BEC) estable en el interior de estrellas compactas, modificando su ecuaci\'on de estado (EdE) y produciendo firmas observables en su enfriamiento y emisi\'on electromagn\'etica.

El objetivo principal de esta tesis es estudiar las propiedades termodin\'amicas de un gas ideal magnetizado de bosones escalares cargados relativistas a bajas temperaturas y campos magn\'eticos arbitrarios, aplicando esto al estudio de los efectos del campo magn\'etico en la estructura de Estrellas de condensado de Bose-Einstein, tomadas estas como una descripci\'on alternativa de las regiones m\'as internas de las ENs. Tambi\'en nos enfocamos en la anisotropía de la ecuaci\'on de estado (EdE), una característica com\'un de los gases cu\'anticos magnetizados compuestos tanto por bosones como por fermiones. Para dar respuesta a este objetivo general, nos proponemos los siguientes objetivos espec\'ificos:

\begin{itemize}
\item Estudiar las propiedades termodin\'amicas, de gases de bosones y fermiones en presencia de un campo magn\'etico uniforme, prestando especial atenci\'on al l\'imite de bajas temperaturas. 
\item Obtener las EdE de un gas mixto \textit{npe} con piones cargados a temperatura cero, propuesto como un modelo del interior de ENs, analizando los efectos del campo magn\'etico en ellas.
\item Resolver un sistema de ecuaciones de estructura anisotr\'opicas que permita la descripci\'on de objetos compactos magnetizados.
\item Estudiar los efectos del campo magn\'etico en la masa y la forma de las estrellas a partir de la relaci\'on masa-radio obtenida con el uso de las ecuaciones de estructura.
\item Obtener otros observables de las estrellas como su deformaci\'on, corrimiento al rojo y momento cuadrupolar de masa.
\end{itemize}

Para abordar estos objetivos, la tesis est\'a organizada de la forma siguiente: el Cap\'itulo \ref{cap1} es introductorio y se dedica al estudio de algunas cuestiones acerca de la estructura interna, la descripci\'on te\'orica de las estrellas de neutrones, estrellas de bosones y de condensado de Bose-Einstein. En el Cap\'itulo \ref{cap2} se investigan las propiedades termodin\'amicas del bosones escalares cargados y fermiones en presencia de un campo magn\'etico y com esto, obtener las ecuaciones de estado del gas mixto magnetizado, analizando el efecto del campo magn\'etico sibre ellas. En el Cap\'itulo \ref{cap3}, por su parte, se estudia el equilibrio hidrodin\'amico de objetos esferoidades a partir de combinar las ecuaciones de Einstein con una m\'etrica axisim\'etrica. Las ecuaciones de estructra all\'i obtenidas se aplican en el Cap\'itulo \ref{cap4} al c\'alculo de las curvas masa-radio y otros observables de las EBE magnetizadas. Al final de la tesis, las conclusiones y recomendaciones presentan un resumen de los principales resultados as\'i como de las posbles direcciones por las cuales creemos que debe continuar la investigaci\'on. Se adjuntan adem\'as dos ap\'endices. El primero contiene las constantes y unidades f\'isicas utilizadas. El segundo explica el procedimiento seguido para la integraci\'on de las ecuaiones de estructura anisotr\'opicas, el cual ha sido eliminado del texto principal a fin de no recargarlo y facilitar su lectura.
    \chapter{Fenomenolog\'ia y descripci\'on te\'orica de las estrellas de neutrones}
\label{cap1}

Este cap\'itulo tiene car\'acter introductorio. En \'el se presenta un conjunto de aspectos relacionados con la fenomenolog\'ia y la modelaci\'on de los Objetos Compactos y Estrellas de Neutrones, todos ellos necesarios para la comprensi\'on de la tesis as\'i como de sus motivaciones.

\section{El interior de una Estrella de Neutrones}

Entre los objetos compactos, las estrellas de neutrones son entornos ideales para investigar estados extremos de la materia, la f\'isica gravitacional y, potencialmente, las interacciones de la materia oscura \cite{DMGrippa2025}. Compuestas principalmente por neutrones densamente empaquetados, las ENs tienen densidades centrales que pueden superar varias veces la densidad de saturaci\'on nuclear ($\rho_{\text{nucl}} \sim 2.7 \times 10^{14} \text{g cm}^{-3}$). Estos reg\'imenes extremos conducen a intensos campos gravitacionales, con gravedad superficial

\begin{equation}\nonumber
g_s = \frac{M}{R^2\sqrt{1-2M/R}} \sim 2 \times  10^{12} \text{ms}^{-2}.
\end{equation}
\noindent para una EN promedio con masa $M\sim 1.4 M_\odot$ y radio $R=10$ km \cite{NSObserations}. Estos campos gravitacionales permiten sondear reg\'imenes f\'isicos inaccesibles en laboratorios terrestres.

El intenso colapso gravitacional que conduce a la formaci\'on de una EN calienta su interior a una temperatura de $\sim$10 MeV $\simeq 10^{11}$ K \cite{Grippa2025}. Sin embargo, las EN se enfr\'ian r\'apidamente con el tiempo mediante la emisi\'on de neutrinos, disminuyendo $T$ en aproximadamente 3-4 \'ordenes de magnitud, raz\'on por la cual las EN se consideran OCs fr\'ios, donde la temperatura, expresada en MeV, es aproximadamente cero. Por ello tomar $T=0$ cuando se modela una EN es, la mayor\'ia de las veces, una buena aproximaci\'on \cite{Schmitt2010}. 

La mayor\'ia de las estrellas de neutrones conocidas tienen campos magn\'eticos. Los valores del campo magn\'etico superficial de las ENs estimados a partir de las observaciones astron\'omicas se encuentran entre los $10^{9}-10^{13}$~G para los p\'ulsares, y en el orden de los $10^{15}$~G para las magnetars\cite{NSObserations}. Para el campo magn\'etico interno puede establecerse una cota m\'axima a partir de comparar la energ\'ia magn\'etica de la estrella con su energ\'ia gravitacional $(4\pi R^3/3)(B_{\text{max}}^2/8\pi) \sim GM^2/R$ \cite{NSObserations}. De esta forma un estimado del campo magn\'etico m\'aximo que puede sostener un OC se obtiene como funci\'on de la masa y el radio de la estrella:

\begin{equation}\nonumber
B_{\text{max}} \sim 2 \times  10^{8} \left( \frac{M}{M_{\odot}} \right) \left( \frac{R}{R_{\odot}} \right)^{-2} \text{G}.
\end{equation}

El origen de los campos magn\'eticos presentes en los entornos astrof\'isicos a\'un no ha podido ser completamente dilucidado\cite{DaryelModelingCO}. En el caso de las ENs, campos magn\'eticos en el orden de hasta $10^{12}$~G pueden ser explicados por la conservaci\'on del flujo magn\'etico de la estrella progenitora durante el colapso que da lugar al OC~\cite{Shapiro1983}. Sin embargo la explicaci\'on de campos magn\'eticos superficiales en el orden de  $10^{15}$ G, como los presentes en las magnetars, precisan de modelos m\'as elaborados. Las propuestas existentes son incapaces de explicar todas las observaciones, por lo que el problema contin\'ua abierto.

Las carater\'isticas generales de las ENs (masa ($M$), radio ($R$), densidad de masa ($\rho$), temperatura ($T$) y campo magn\'etico superficial ($B_s$) e interno ($B_i$)) se resumen en la Tabla \ref{tablaEN}.

\begin{table}[ht]
	\centering
    \caption{Valores t\'ipicos de magnitudes que caracterizan las ENs \cite{DMGrippa2025,camenzind2007compact,Shapiro1983, NSObserations}}
	\begin{tabular}{|c|c|c|c|c|r|}
		\hline
		$M$($M_{\odot}$) &
		$R$(km)&
		$\rho (\text{g}/\text{cm}^3)$ &
		$T$(K) &
		$B_s\text{(G)}$&
		$B_i\text{(G)}$ \\
		\hline
		$\sim 1.2 - 2.5$ &
		$\sim 10 - 15$&
		$\sim 10^7$-$10^{15}$ &
		$\sim 10^5$-$10^{11}$ &
	    $10^{9}$-$10^{15}$ &
		$\lesssim 10^{18}$ \cr
		\hline
	\end{tabular}
\vspace{10pt}
	\label{tablaEN}
\end{table}
La estructura interna de las ENs es t\'ipicamente modelada en capas, como muestra la Figura \ref{NSstructure}. Basados en nuestro entendimiento actual, tenemos las siguientes regiones\cite{DMThesis}:
\begin{figure}[!ht]
	\centering
	\includegraphics[width=0.6\linewidth]{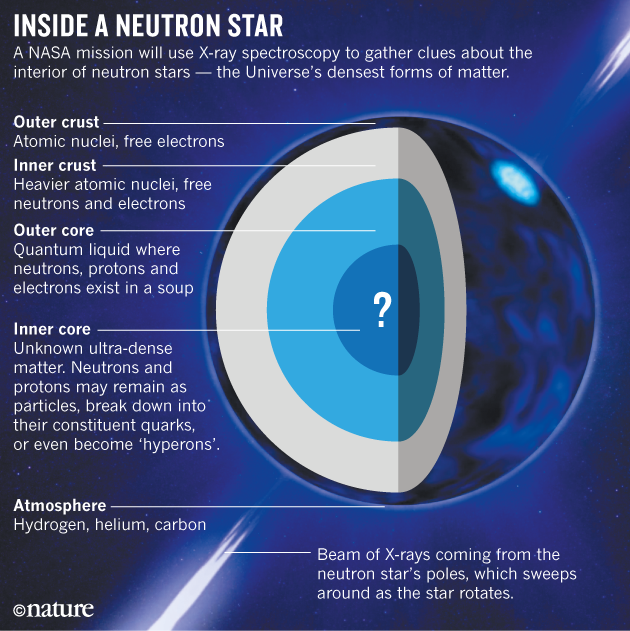}
	\caption{ Ilustraci\'on representativa de la estructura de una estrella de neutrones, publicada en Nature\cite{Gibney2017} para conmemorar el inicio de la misi\'on NICER de la NASA.}
\end{figure}\label{NSstructure}
\begin{itemize}
    \item La atm\'osfera de la EN es una delgada capa de plasma responsable de formar el espectro de su radiaci\'on electromagn\'etica t\'ermica. Esta capa, de solo unos pocos mil\'imetros de espesor, contribuye insignificativamente a la masa total de la estrella, pero contiene valiosa informaci\'on sobre los par\'ametros superficiales (temperatura efectiva, gravedad, composici\'on qu\'imica, intensidad y geometr\'ia del campo magn\'etico superficial).
    
	\item La envoltura, que es una capa s\'olida formada por una red cristalina de hierro de alrededor de unos cientos de metros.
	
	\item  La corteza externa es una regi\'on con un espesor que var\'ia entre 0.3 y 0.5 km, compuesta de materia con una densidad inferior a $\rho_{\text{drip}} \approx 4 \times 10^{11}$ g/cm$^3$. A esta densidad, la materia consiste en una red de Coulomb de n\'ucleos inmersos en un gas de electrones.
    
    \item La corteza interna abarca aproximadamente entre 1 y 2 km \cite{Piekarewicz2022} y presenta una densidad que va desde $\rho_{\text{drip}}$ hasta aproximadamente la mitad de la densidad de saturaci\'on $\rho_0 = 2.8 \times 10^{14}$ g/cm$^3$ (o $\rho_0 = 0.15$ fm$^{-3}$). En esta regi\'on, n\'ucleos altamente enriquecidos en neutrones est\'an inmersos en un gas superfluido de neutrones. Debido al limitado entendimiento de la interacci\'on nucle\'on-nucle\'on, se requieren enfoques te\'oricos basados en la teor\'ia de muchos cuerpos.
    
	\item El n\'ucleo externo cubre un rango de densidad de $0.5\rho_0 \leq \rho \leq 2\rho_0$ y tiene un espesor de varios kil\'ometros. Consiste principalmente en neutrones, con una fracci\'on de protones y electrones. El estado de dicha materia se determina por las condiciones de neutralidad de carga y equilibrio beta, suplementado por un modelo microsc\'opico de la interacci\'on entre los nucleones. 
    
    \item El n\'ucleo interno, donde $\rho \gtrsim 2 \, \rho_0$, ocupa la regi\'on central de la estrella. Su densidad central puede ser de hasta $(10-15) \, \rho_0$\cite{Piekarewicz2022}. Su composici\'on es hasta el momento incierta debido a la falta de aproximaciones te\'oricas y experimentales para el r\'egimen de densidades t\'ipicas de esta regi\'on. Las Ecuaciones de Estado (EdE) dependen del modelo utilizado. Muchas hip\'otesis han sido formuladas, las cuales predicen la aparici\'on de nuevos fermiones y/o condensados de bosones. Las cuatro principales son: la aparici\'on de hiperones, principalmente los $\Lambda$ y $\Sigma$, condensados de piones, condensados de kaones o una transici\'on de fase a materia de quarks compuesta de quarks deconfinados $u$, $d$ y $s$ con una pequeña fracci\'on de electrones o no.
\end{itemize}

\section{Modelaci\'on de estrellas de neutrones}

De manera general, cuando se habla de uno u otro modelo de ENs, a lo que se hace referencia es a distintas combinaciones de part\'iculas, interacciones y fases que podr\'ian coexistir en los n\'ucleos de estos objetos. Para saber si un modelo de EN es o no gravitacionalmente estable, lo usual es combinar sus ecuaciones de estado con las llamadas ecuaciones de estructura. Las EdE est\'an relacionadas con la f\'isica micr\'oscopica de la estrella y contienen toda la informaci\'on de la materia y los campos que la componen, as\'i como de las interacciones entre ellos. Las ecuaciones de estructura, en cambio, est\'an relacionadas con la f\'isica macrosc\'opica del objeto compacto. Ellas son consecuencia de las ecuaciones de Einstein y expresan el equilibrio hidrodin\'amico que se establece en el interior de la estrella entre la fuerza de gravedad y la presi\'on que ejerce la materia \cite{Shapiro1983,camenzind2007compact}. La relaci\'on entre las EdE y las ecuaciones de estructura viene dada a trav\'es de la presi\'on y densidad de energ\'ia  de la materia que compone la estrella.

Los modelos de ENs que hasta el momento se consideran m\'as realistas est\'an basados en la f\'isica nuclear experimental y en c\'alculos relativamente precisos de muchos cuerpos para modelos de materia nuclear densa \cite{camenzind2007compact,NSObserations}. No obstante, todos ellos est\'a limitados por la imposibilidad actual de alcanzar densidades superiores a la densidad nuclear en el laboratorio \cite{NSObserations}. Esto significa que la descripci\'on de la materia y sus interacciones en reg\'imenes en los cuales $\rho>\rho_{nuc}$ se hace siempre de manera aproximada, a partir de extrapolar los comportamientos conocidos para densidades menores \cite{Schmitt2010}. Estas limitaciones dan lugar a una larga lista de conjeturas sobre lo que ocurre en el interior de una EN. Hasta el momento resulta casi imposible saber cu\'ales llegan a producirse y cu\'ales no \cite{NSObserations}. Ello se debe, de un lado a las limitaciones observacionales, y del otro al hecho de que la mayor\'ia de las EdE propuestas cumplen con las de restricciones te\'oricas generales que pueden imponerse sobre las masas y radios de los OCs a fin de acotar sus  valores posibles. Dichas rectricciones son \cite{NSObserations,Schmitt2010}:

\begin{itemize}
	\item Requerimiento de estabilidad gravitacional: el radio $R$ de un OC de masa $M$ debe ser mayor que el radio de Schwarzschild $R_s = 2 G M$, donde $G = 6.711 \times 10^{-45}$~MeV$^{-2}$ es la constante de gravitaci\'on. Para $R<R_s$ la estrella colapsa formando un AN \cite{Schmitt2010}.
	\item Requerimiento de presi\'on finita: suponiendo que la estrella es un objeto esf\'erico, las Ecuaciones de Einstein requieren que $R >9 /4 G M$ \cite{Shapiro1983}.
	\item Requerimiento de causalidad: exige a las EdE de que la velocidad del sonido en la estrella no supere a la velocidad de la luz, lo cual implica que $R > 2.9 G M$ \cite{NSObserations}.
\end{itemize}
Adem\'as de las restricciones te\'oricas ya mencionadas, las observaciones astron\'omicas aportan constantemente nuevas cotas para las masas y los radios de las ENs. La \'ultima d\'ecada ha presenciado grandes avances en las observaciones de ENs, gracias al descubrimiento de ENs s\'uper masivas, mejoras cualitativas en el entendimiento de incertidumbres asociadas a las mediciones de radios con rayos-X \cite{Bogdanov2021}. 

La detecci\'on de ondas gravitacionales ha generado una serie de posibilidades observacionales que se espera devengan en los pr\'oximos años en una nueva forma de explorar las ENs y el Universo, as\'i como la detecci\'on por LIGO-Virgo de la OG GW170817 originada de la fusi\'on de dos ENs \cite{abbott2017gw170817}. Por otra parte, a partir de observatorios como el \textit{Neutron star Interior Composition Explorer (NICER)}, construidos espec\'ificamente para recolectar informaci\'on acerca de la composici\'on de las ENs, se espera encontrar evidencias de la existencia en su interior de nuevas fases y tipos de materia. Las estrellas de neutrones pueden emitir ondas gravitacionales si experimentan alg\'un tipo de deformaci\'on, ya sea por perturbaciones inducidas por una compañera en un sistema binario, por rotaci\'on o por campos magn\'eticos que las desv\'ian de la esfericidad perfecta\cite{Samantha2024}. Hasta ahora, determinar con alta precisi\'on los par\'ametros de estas estrellas no es posible con la generaci\'on actual de detectores, algo que se espera superar en el futuro cercano\cite{ligo2023}. Las estrellas aisladas magnetizadas tambi\'en sufren deformaciones, convirti\'endose en esferoides con un cuadrupolo de masa asociado, lo que las convierte en otra fuente de ondas gravitacionales. La comparaci\'on entre modelos te\'oricos y observaciones es fundamental para avanzar en este campo.

\begin{figure}[!ht]
	\centering
	\includegraphics[width=0.8\linewidth]{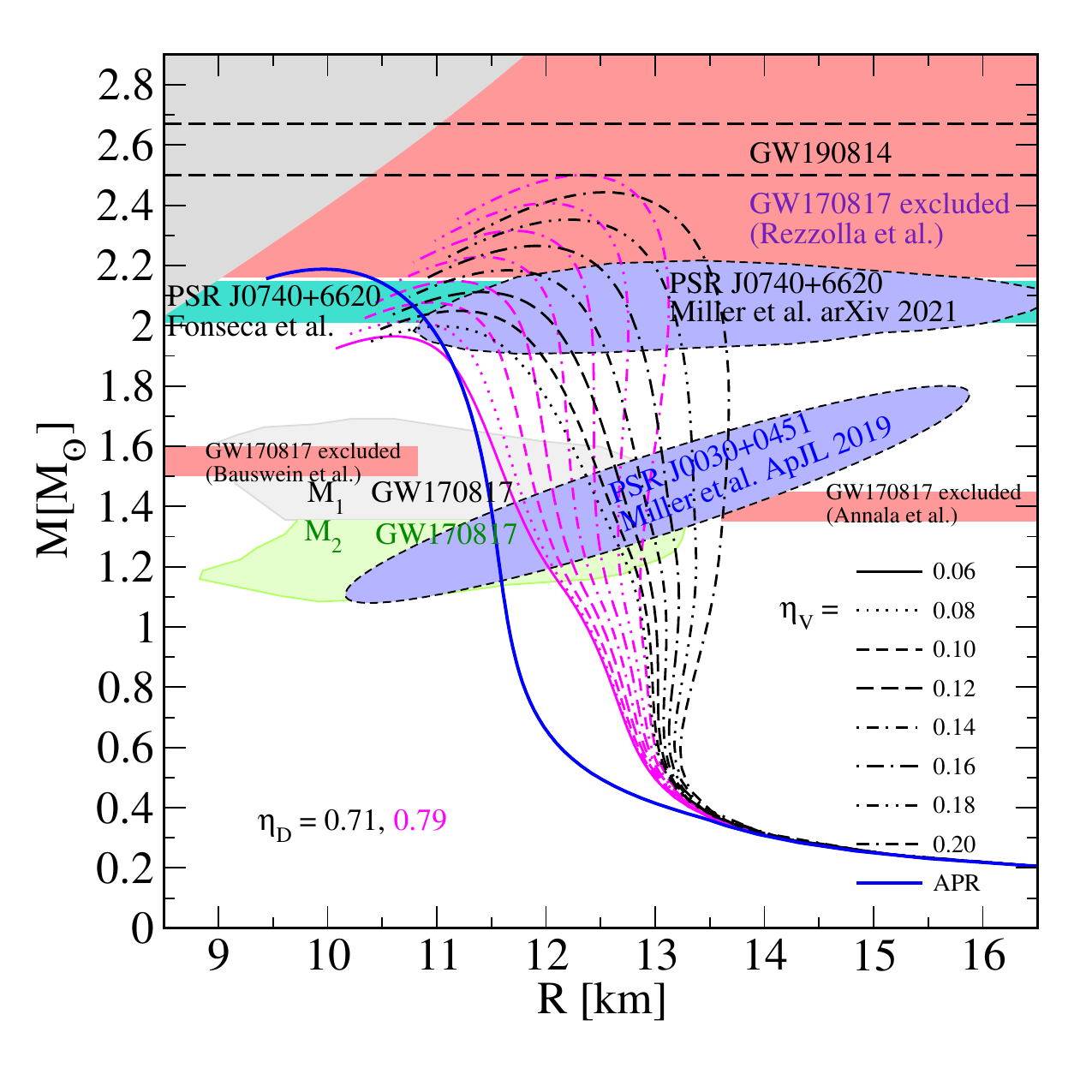}
	\caption{Relaciones masa-radio para modelos de EdE con algunas restricciones observacionales \cite{Ayriyan2021}.}\label{MRNICER}
\end{figure}
Actualmente, las mediciones m\'as confiables han resultado en valores observacionales para las masas de las ENs alrededor de las 2 M$_\odot$. Estas observaciones han hecho que dentro de la comunidad astrof\'isica haya una b\'usqueda intensa de modelos de ENs cuyas EdE produzcan estrellas con masas m\'aximas M> 2 M$_\odot$ \cite{Masquerade}. Sin embargo, las EdE que producen estrellas de masas menores siguen siendo importantes para la descripci\'on de algunas de las capas internas de la estrella (como en una Estrella H\'ibrida) o para explicar observaciones que no respondan a las caracter\'isticas de las ENs can\'onicas.


En la Figura \ref{MRNICER} se muestran varias secuencias de estrellas compactas dadas por diferentes EdE \cite{Ayriyan2021}. Las regiones coloreadas corresponden a mediciones de p\'ulsares o zonas prohibidas que sirven como restricciones para la EdE de estrellas compactas. La banda verde (por encima de 2M$_{\odot}$) corresponde a la medici\'on actualizada de la masa de PSR J0740+6620, recientemente mejorada a una medici\'on masa-radio por NICER, representada por la regi\'on elipsoidal azul (resultado del equipo Maryland-Illinois). El otro elipse azul corresponde a la medici\'on de masa y radio de PSR J0030+0451 por NICER, mientras que las zonas gris y verde claro representan las estimaciones de las componentes $M_1$ y $M_2$ del sistema binario de la fusi\'on GW170817. Las bandas rojas son regiones excluidas derivadas de observaciones de GW170817. Las l\'ineas horizontales negras discontinuas indican los l\'imites superior e inferior para la masa ($2.59^{+0.08}{-0.09}~M_{\odot}$) del componente m\'as ligero en el evento de fusi\'on binaria GW190814.
 
\section{Estrellas de bosones}

Si los objetos compactos est\'an compuestos por algo distinto a la materia bari\'onica ordinaria, se dice que son ex\'oticos. Un ejemplo de estos son las estrellas de bosones, que son objetos hipot\'eticos compuestos por bosones, cuya estabilidad surge del equilibrio entre la atracci\'on gravitacional y la naturaleza dispersiva de un campo escalar. Fueron propuestas por primera vez a finales de la d\'ecada de 1960 por Kaup \cite{Kaup1968} y Ruffini y Bonazzola \cite{Ruffini1969}, quienes sentaron las bases te\'oricas para su estudio. Desde entonces, estas configuraciones han sido investigadas activamente durante m\'as de medio siglo \cite{Liebling2023, Chavanis2012}, revelando propiedades \'unicas que las distinguen de otros objetos compactos.

A diferencia de las estrellas de neutrones, sostenidas por la presi\'on de degeneraci\'on fermi\'onica, las estrellas de bosones no requieren una ecuaci\'on de estado cl\'asica. En su lugar, su estabilidad emerge como una soluci\'on autoconsistente del acoplamiento entre un campo escalar complejo y la gravedad. Ruffini y Bonazzola \cite{Ruffini1969} demostraron que este sistema puede formar estructuras compactas con masa finita y un perfil radial bien definido, donde los efectos cu\'anticos del campo bos\'onico act\'uan como una ``presi\'on'' estabilizadora. Esta soluci\'on constituye la primera descripci\'on rigurosa de lo que hoy conocemos como estrella de bosones, mostrando que tales objetos son compatibles con la relatividad general.


Los primeros estudios de Ruffini y Bonazzola \cite{Ruffini1969} sobre estrellas de bosones revelaron una limitaci\'on fundamental: en el caso m\'as simple (un campo escalar libre sin autointeracciones), la masa m\'axima de estas configuraciones era extremadamente pequeña, del orden de $M_{max} \sim m_{Pl}^2/m$, donde $m$ es la masa del bos\'on y $m_{Pl}$ es la masa de Planck. Para bosones ligeros, como por ejemplo piones, con masas $m_{\pi} = 140 MeV$ esto implica masas m\'aximas de apenas $10^{-20}$ $M_{\odot}$, descartando su viabilidad como objetos astrof\'isicos observables.

Este resultado inicial plante\'o un desaf\'io te\'orico, pero tambi\'en una posible soluci\'on: la introducci\'on de autointeracciones no lineales en el campo escalar. Cuando se incluyen t\'erminos de autointeracci\'on atractivos o repulsivos en el potencial del campo, la masa m\'axima de la estrella de bosones puede aumentar significativamente. en particular, para acoplamientos suficientemente grandes, las estimaciones muestran que $M_{max}$ puede alcanzar valores de inter\'es astrof\'isico\cite{Chavanis2012, Latifah2014Bosons}.


\subsection{Estrellas de Condensado de Bose-Einstein}

En años recientes, se ha observado la condensaci\'on de Bose-Einstein en vapores diluidos de elementos alcalinos atrapados magn\'eticamente \cite{Chavanis2012}. En este caso, la condensaci\'on de Bose-Einstein se realiza en una trampa artificial. En la naturaleza, existe una trampa natural: la gravitatoria. Aunque hasta ahora no hay observaciones astron\'omicas que confirmen la existencia de una estrella de Bose fr\'ia compuesta de un condensado diluido de Bose-Einstein, es de gran inter\'es acad\'emico estudiar este tipo de estrella. Recientemente, se ha demostrado que configuraciones particulares de haces l\'aser intensos y fuera de resonancia pueden inducir una interacci\'on de tipo gravitatorio entre \'atomos ubicados dentro de la longitud de onda del l\'aser \cite{Chavanis2016}. 
A partir de estos estudios y del \'exito en el ajuste de la curva de enfriamiento de Cassiopea A, que la presi\'on resultante de la interacci\'on repulsiva vino a jugar el papel principal en la construcci\'on de modelos de estrellas de bosones \cite{Chavanis2012,Latifah2014Bosons}. En estos modelos relativamente recientes, conocidos en la literatura como Estrellas de condensado de Bose-Einstein, la gravedad es balanceada por la presi\'on que resulta de las fuerzas repulsivas, y el esquema de combinar EdE con ecuaciones de estructura es retomado para la b\'usqueda de las configuraciones estelares estables.

Las EBE se proponen como una alternativa para la descripci\'on del n\'ucleo de una EN. En una estrella de fermiones, el empuje de la gravedad es contrarrestrado en \'ultima instancia por la presi\'on del gas degenerado o presi\'on de Pauli. Dadas las altas densidades que existen en el interior del los OCs, incluso en el caso de que el gas de fermiones se suponga ideal y a temperatura cero, la presi\'on de Pauli es lo suficientemente alta para balancear la gravedad.
Por el contrario, un gas ideal de bosones a temperatura cero no ejerce presi\'on porque todas las part\'iculas est\'an en el condensado de Bose-Einstein \cite{PathriaBeale2021}. Pero a\'un en este caso el colapso gravitacional puede evitarse gracias a la llamada presi\'on de Heisenberg, es decir, al hecho de que un gas bos\'onico condensado no puede ser comprimido infinitamente porque, debido al Principio de Incertidumbre:
\begin{equation}\label{PI}
\Delta p \Delta x \thickapprox \hbar, 
\end{equation}
\noindent a medida que la incertidumbre en el volumen decrece, la del momentum aumenta, aumentando con ella la presi\'on. Los par\'ametros que definen el tama\~no y la masa de la estrella resultante son la masa de los bosones y la fortaleza de la repulsi\'on entre ellos. Y precisamente parte del \'exito  de estos modelos se debe a que es posible, a trav\'es de una selecci\'on apropiada de la masa y la fortaleza de la interacci\'on, lograr con \'el estrellas con masas del orden de dos masas solares \cite{Chavanis2012}.

Muchas teor\'ias de part\'iculas predicen que los bosones d\'ebilmente interactuantes son abundantes en el Universo y podr\'ian haber desempeñado un papel significativo en la evoluci\'on del Universo temprano, por lo que se ha sugerido que estas EBE podr\'ian ser la materia oscura del Universo, lo que las convertir\'ia en la masa faltante cosmol\'ogica.
    \chapter{Ecuaciones de estado de un gas magnetizado de bosones escalares cargados y fermiones.}\label{cap2}
    
 En este cap\'itulo obtendremos las propiedades termodin\'amicas de gases de bosones escalares cargados (piones), as\'i como las de fermiones cargados (electrones y protones) y neutros (neutrones), en presencia de un campo magn\'etico uniforme. Dedicaremos especial atenci\'on al  l\'imite de bajas temperaturas, para luego obtener las ecuaciones de estado  a a $T = 0$ de un gas de piones inmerso en materia nuclear (\textit{npe}). Estas ecuaciones se utilizar\'an en el siguiente cap\'itulo para obtener los observables astrof\'isicos de estrellas esferoidales magnetizadas. Parte de los resultados aqu\'i presentados son contribuciones originales del autor.

 \section{Potencial termodin\'amico de gases de fermiones y bosones en presencia de campo magn\'etico}

Gases de  bosones y fermiones en presencia de un campo magn\'etico uniforme y constante en presencia de un campo magn\'etico en la direcci\'on $z$ sus espectros de energ\'ias y  energ\'ias del estado b\'asico  vienen  vienen dadas por:

    \begin{equation} \label{FermionsGS}
      \varepsilon_F = 
      \left\{ 
      \begin{array}{ll}
        \sqrt{p_3^2 + m_F^2 + 2eBl}, & \text{Fermiones cargados} \\
        \sqrt{p_3^2 + \left(\sqrt{p_{\bot}^2 + m_F^2} + \eta \kappa B \right)^2}, & \text{Fermiones neutros}
      \end{array}
      \right.
    \end{equation}
    
    \begin{equation} \label{bosonsGS}
      \varepsilon_b = 
      \left\{
      \begin{array}{ll}
        \sqrt{p_3^2 + m_B^2 + 2eB(l+1/2)}, & \text{Bosones escalares cargados} \\
        \sqrt{p_3^2 + p_{\bot}^2 + 2\eta \kappa B \sqrt{p_{\bot}^2 + m^2}}, & \text{Bosones vectoriales neutros}
      \end{array}
      \right.
    \end{equation}
    
    En las Ecs.(\ref{FermionsGS}) y (\ref{bosonsGS}), $m_B$ y $m_F$ denotan las masas del bos\'on y el fermi\'on; $q$ es la carga el\'ectrica $e$ de las  part\'iculasy $\kappa =-1.913\mu_N$ es el momento magn\'etico an\'omalo del neutr\'on, ocn $\mu_N$ siendo en magnet\'on nuclear. 
    
     En el marco del formalismo de tiempo imaginario \cite{Amanda}, la expresi\'on para el  potencial termodin\'amico es:

    \begin{equation}
    \Omega_{F,B}(T, \mu, B) = -eB\,T \sum_{p_4} \sum_{l=0}^{\infty} \alpha_l \int \frac{dp_3}{2\pi} \,  
    \text{Tr} \ln D^{-1}_l(p^*),
    \end{equation}\label{OmegaFermiBoson}
    
    \noindent donde $p^{*}=ip_4-\mu$, y la suma por $p_4$ es la suma por los momenta de Matsubara \cite{Ayala2012}.   $\alpha_l = 2 - \delta_{l0}$  la cuenta de la degeneraci\'on  de spin y $D^{-1}_l(p^*)$ es el inverso del propagador de las part\'iculas en presencia de un campo magn\'etico. La expresi\'on expl\'icita del propagador tiene la forma

    \begin{equation}
    D_l^{-1}(p^*) = 
    \begin{cases}
    \left( (i\omega_n - \mu)^2 + p_3^2 + (2l + 1)eB + m^2 \right), & \text{Bosones} \\
    \left( (i\omega_n - \mu)^2 + p_3^2 + 2eBl + m^2 \right), & \text{Fermiones}
    \end{cases}
    \end{equation}
    
Una vez realizada la suma sobre las frecuencias de Matsubara \cite{Amanda}, el  potencial termodin\'amico puede escribirse en dos contribuciones: la conocida contribuci\'on del \textit{vacío} $\Omega_V(B)$, que depende \'unicamente de la intensidad del campo magn\'etico y requiere renormalizaci\'on; y la llamada contribuci\'on \textit{estadística} $\Omega_{\text{st}}(T, \mu, B)$, que depende de la temperatura, el potencial químico y la intensidad del campo magn\'etico, obteniendo:
	
	\begin{align}\label{desglose}
		\Omega(T, \mu, B) &= \Omega_V(B) + \Omega_{\text{st}}(T, \mu, B) \nonumber \\
		&= \frac{eB}{4\pi^2} \sum_{l=0}^{\infty} \alpha_l \int_{-\infty}^{\infty} E_l(p_3) dp_3 \nonumber \\
		&\quad + \frac{eB}{4\pi^2 \beta} \sum_{l=0}^{\infty}\alpha_l\int_{-\infty}^{\infty} dp_3 \ln \left[ (1 + \eta e^{-\beta(E_l(p_3) - \mu)})(1 + \eta e^{-\beta(E_l(p_3) + \mu)}) \right],
	\end{align}
	
\noindent donde $\eta=\pm 1$ para fermiones 1 y para bosones $-1$, $\beta = 1/T$ es el inverso de la temperatura absoluta $T$, $\mu$ es  el potencial químico de las part\'iculas; $l$ representa la cuantizaci\'on de los niveles de Landau para el momento transversal al campo magn\'etico. 

 La parte estad\'istica puede dividirse en la contribuci\'on del nivel de Landau m\'as bajo (LLL, por sus siglas en ingl\'es) y la de los estados excitados.

La suma sobre los niveles de Landau y la integral sobre $p_3$ en la contribuci\'on del vacío pueden realizarse directamente mediante el m\'etodo de tiempo propio de Schwinger, pero dado que esta contiene tanto la contribuci\'on como la polarizaci\'on del vacío, ambas deben aislarse y regularizarse \cite{19Schwinger1951}. Así, la contribuci\'on finita al potencial termodin\'amico del vacío se expresa como:
	
    \begin{equation}\label{Pot_Vacio}
    \Omega_V = 
    \begin{cases} 
    -\frac{eB}{16\pi^2} \int_0^{\infty} \frac{ds}{s^2} \left[\frac{1}{\sinh(eBs)} - 1 + \frac{eBs}{6}\right]e^{-m^2 s} & \text{Bosones cargados} \\[10pt]  
    \frac{eB}{8\pi^2} \int_0^\infty \frac{ds}{s^2} \left[ \coth(eBs) - \frac{1}{eBs} - \frac{eBs}{3} \right] e^{-m_F^2 s} & \text{Fermiones cargados} \\[10pt]
    \frac{1}{8\pi^2} \int_0^{\infty} \frac{ds}{s^3} \left[\cosh\left(\kappa B s\right) -1\right] e^{-m_F^2s} & \text{Fermiones neutros}
    \end{cases}
    \end{equation}

En los estudios termodin\'amicos de las propiedades del  vacío regularizado suelen ignorarse. Sin embargo, este t\'ermino contribuye a la energía del vacío $\Omega_V^R(B)$ como energía de punto cero, que desempeña un papel crucial en las propiedades magn\'eticas del gas y, en consecuencia, en la ecuaci\'on de estado como veremos mas adelante.

Consideremos ahora la parte estadística del potencial termodin\'amico dada por la Ec. (\ref{desglose}), para valores arbitrarios del campo magn\'etico, a temperatura y densidad finitas. Si desarrollamos en serie de Taylor el logaritmo, y usamos la identidad
	
	\begin{equation}
		e^{-\beta n\sqrt{p_3^2+2eB(l+1/2)+m^2}} = \frac{\beta n}{2}\int_{0}^{\infty}\frac{ds}{\sqrt{\pi}s^{3/2}}e^{-\frac{\beta^2 n^2}{4s}-s(p_3^2+2eB(l+1/2)+m^2)},
	\end{equation}
	
\noindent la integral sobre $p_3$ y la suma por $l$ pueden realizarse directamente, obteniendo \cite{Ayala2012}:
	
\begin{equation}\label{Pot_st}
\Omega_{st}(T,\mu,B) = 
\left\{
\begin{array}{ll}
-\dfrac{eB}{4\pi^2} \int_{0}^{\infty} \dfrac{ds}{s^2} e^{-(m_B^2+eB)s} \left(1 + \dfrac{1}{e^{2eBs}-1}\right) \displaystyle \sum_{k=1}^{\infty} e^{-\frac{k^2 \beta^2}{4s}} \cosh(\mu \beta k), & \text{Bosones} \\
\\
\frac{eB}{4\pi^2} \int_{0}^{\infty} \dfrac{ds}{s^2} e^{-m_F^2 s} \left(1 - \dfrac{2}{1 - e^{-2eBs}}\right) \displaystyle \sum_{k=1}^{\infty} (-1)^k e^{-\frac{k^2 \beta^2}{4s}} \cosh(\mu \beta k), & \text{Fermiones}
\end{array}
\right.
\end{equation}

El primer t\'ermino dentro del parent\'esis es la contribuci\'on del LLL, $\Omega_{LLL}$, mientras que el segundo  se  corresponde con  la contribuci\'on de los niveles de Landau excitados  $\Omega_{l\neq 0}$, Estas expresiones son v\'alidas para valores arbitrarios del campo magn\'etico y la temperatura.

\section{L\'imite de bajas temperaturas.}
En esta secci\'on nos enfocaremos en determinar las propiedades termodin\'amicas en el l\'imite de bajas temperaturas, tanto para bosones como fermiones.

\subsection{Bosones}

Comencemos por calcular el potencial  termodin\'amico para los bosones en el  r\'egimen de baja temperatura, caracterizado por temperaturas muy por debajo de la energía del estado fundamental ($T \ll m_B$).
	
Para explorar las propiedades termodin\'amicas en este r\'egimen, recurriremos al m\'etodo del descenso m\'as pronunciado (\textit{steepest descent}) \cite{Amanda} que nos permite aproximar la integral sobre $s$ en las Ecs. (9a)-(9c), en el límite $m_B/T \to \infty$, de la siguiente manera:
	
	\begin{equation}
		\int_{0}^{\infty} ds\, g(s)e^{\frac{m_b}{t}f(s)} \sim \frac{\sqrt{2\pi t}\, g(s_0)}{|m_b f''(s_0)|^{1/2}}e^{\frac{m_b}{t}f(s_0)},
	\end{equation}

\noindent donde $s_0$ es un punto m\'aximo de la funci\'on $f(s)$, que en nuestro caso es $f(s) = -s - \frac{n^2}{4s}$ y $s_0 = n/2$.
Introduzcamos  variables adimensionales: campo magn\'etico $b = B/B_c^{ch}$ (siendo $B_c^{ch} = m^2/e$ el campo magn\'etico cr\'itico de part\'iculas cargadas), energía del estado b\'asico $m_b \equiv m_B/m = \sqrt{1+b}$, temperatura $t \equiv T/m$ y el potencial químico como $x \equiv \mu/m$,

El potencial termodin\'amico adimensional $\hat{\Omega} \equiv \Omega^{LT}/m^4$, en el r\'egimen de baja temperatura tiene la forma 

$$\hat{\Omega}_{st}^{LT} = \hat{\Omega}_{LLL}^{LT}(t,x,b) + \Omega_{l\neq 0}^{LT}(t,x,b),$$
\noindent cuya forma expl\'icita queda como
	
	\begin{align}
		\hat{\Omega}_{LLL}^{LT}(t,x,b) &= -b m_b^{1/2} \left(\frac{t}{2\pi}\right)^{3/2} \text{Li}_{3/2}(z), \\
		\hat{\Omega}_{l\neq 0}^{LT}(t,x,b) &= -b m_b^{1/2} \left(\frac{t}{2\pi}\right)^{3/2} \sum_{k=1}^{\infty} \frac{z^k}{k^{3/2}} \frac{1}{e^{k\gamma} - 1}, \label{Omegab}
	\end{align}
	
\noindent  siendo $\text{Li}_s(z)$ la funci\'on polilogarítmica\footnote{En una representaci\'on en serie $\text{Li}_s(z) = \sum_{k=1}^{\infty} \frac{z^k}{k^s}$.} de orden $s$; $z \equiv e^{(x - m_b)/t}$, siendo $z$ la fugacidad y $\gamma = \frac{b}{t m_b}$ es el campo magn\'etico escalado.

A partir del potencial termodin\'amico, podemos calcular todas las propiedades termodin\'amicas del sistema. 
    
La densidad de part\'iculas adimensional $\hat{N} = \frac{N}{m^3} = -\left(\frac{\partial \hat{\Omega}}{\partial x}\right)_{t,b}$ toma la forma expl\'icita
	\begin{align}
		\hat{N}_{LLL}(t,x,b) &= \frac{m_b^{1/2} t^{1/2} b}{(2\pi)^{3/2}} \text{Li}_{1/2}(z) \\
		\hat{N}_{l\neq 0}^{LT}(t,x,b) &= \frac{m_b^{1/2} t^{1/2} b}{(2\pi)^{3/2}} \sum_{k=1}^{\infty} \frac{z^k}{k^{1/2}} \frac{1}{e^{k\gamma} - 1}.
	\end{align}
	
$\hat{N}_{LLL}$  diverge cuando $z\to 1$ alcanz\'andose el condensado de Bose Eisntein en presencia de campo magn\'etico \cite{QuinteroAngulo2022}. 
La densidad de energía estadística se define como $\hat{E}_{st} = \hat{\Omega}_{st} + t \hat{S} + x \hat{N}$, con la densidad de entropía definida como $\hat{S} = -\left(\frac{\partial \hat{\Omega}}{\partial t}\right)_{x,b}$, obteni\'endose:
	\begin{align}
		\hat{E}_{LLL}(t,x,b) &= \frac{m_b^{1/2} b t^{1/2}}{4\sqrt{2}\pi^{3/2}} \left[2m_b \text{Li}_{1/2}(z) + t \text{Li}_{3/2}(z)\right], \label{EnerLLL}\\
		\hat{E}_{l\neq 0}(t,x,b) &= \sum_{k=1}^{\infty} \frac{\sqrt{t} b z^k}{(2\pi k)^{3/2} \sqrt{m_b} (e^{k\gamma} - 1)^2} \left[ -k m_b^2 + (1 + 2b) k e^{n\gamma} \right.\left. + \frac{1}{2} m_b t (e^{k\gamma} - 1) \right].
	\end{align}
Definimos adem\'as la magnetizaci\'on adimensional $\hat{\mathcal{M}} = -\left(\frac{\partial \hat{\Omega}}{\partial b}\right)_{t,x}$ \footnote{Para recuperar las dimensiones físicas, debe multiplicarse por el factor $e/m$}.
La magnetizaci\'on puede descomponerse en contribuciones estadísticas y del vacío ($\hat{\mathcal{M}}(t,x,b) = \hat{\mathcal{M}}_{st}(t,x,b) + \hat{\mathcal{M}}_{vac}(b)$). La contribuci\'on del vacío est\'a dada por:
	\begin{align}
		\hat{\mathcal{M}}_{vac}(b) &= -\frac{b}{2\pi^2} \zeta^{(1,0)} \left(-1, \frac{b+1}{2b}\right) + \frac{1}{8\pi^2} \zeta^{(1,1)} \left(-1, \frac{b+1}{2b}\right) \nonumber \\
		&\quad - \frac{b}{96\pi^2} - \frac{1}{32\pi^2 b} + \frac{b \log(4)}{96\pi^2} + \frac{b \log(b)}{48\pi^2}
	\end{align}
	
	y es siempre positiva, lo que indica que el vacío exhibe un comportamiento paramagn\'etico. Adem\'as, es una funci\'on mon\'otonamente creciente de la intensidad del campo magn\'etico.
	
	La magnetizaci\'on estadística se expresa como 
	\begin{align}
        \hat{\mathcal{M}}_{st}(t,x,b) &= \hat{\mathcal{M}}_{LLL}(t,x,b) + \hat{\mathcal{M}}_{l\neq 0}(t,x,b) \\
		\hat{\mathcal{M}}_{LLL}(t,x,b) &= \frac{\sqrt{t}}{4(2\pi m_b)^{3/2}} \left(-2b m_b \text{Li}_{1/2}(z) + (4 + 5b) t \text{Li}_{3/2}(z)\right),\label{magneLLL}\\
		\hat{\mathcal{M}}_{l\neq 0}(t,x,b) &= \sum_{k=1}^{\infty} \frac{z^k}{4} \left(\frac{t}{2\pi k m_b}\right)^{3/2} \frac{-4 + e^{k\gamma}(-6k\gamma + 4) + 2k\gamma - b[5 + e^{k\gamma}(4k\gamma - 5) - 2k\gamma]}{(e^{k\gamma} - 1)^2}.
	\end{align}
		
	Para $t \to 0$ a $\hat{N}$ fijo tenemos:
	
	\begin{equation}
		\hat{\mathcal{M}}_{st}(t \to 0,x,b) = \lim_{t \to 0} \hat{\mathcal{M}}_{LLL}(t,x,b) = -\frac{\hat{N}}{2\sqrt{1 + b}},
	\end{equation}
	
\noindent que es siempre negativa. 
y se puede escribir como 
$$\mathcal{M}_{st} = -\frac{e N}{2 m_B}$$ que para $b\to 0$ da una magnetizaci\'on diferente de cero mostrando que el sistema se vuelve ferro-diamagn\'etico \cite{21Rojas1996}.

\subsection{Fermiones}

    Siguiendo un procedimiento an\'alogo al utilizado para los bosones, en el r\'egimen de bajas temperaturas, la contribuci\'on t\'ermica para fermiones cargados puede expresarse como:
    
    \begin{equation}
    \Omega_{F}^{\text{LLL}} = -\frac{eB T^2}{2\pi} \left[ \text{Li}_2(-e^{(\mu_F - m_F)/T}) + \text{Li}_2(-e^{-(\mu_F + m_F)/T}) \right],
    \end{equation}
    
    Cuando $\mu \approx \varepsilon_F$ (energía de Fermi):
    
    \begin{equation}
    \Omega_{F}^{\text{LLL}} \approx -\frac{eB}{4\pi^2} \left[ \mu_F \sqrt{\mu_F^2 - m_F^2} + \frac{\pi^2 T^2}{3} \frac{\mu_F}{\sqrt{\mu_F^2 - m_F^2}} \right].
    \end{equation}

    En este r\'egimen, la contribuci\'on dominante proviene de los niveles cercanos al potencial químico. Usando el m\'etodo del descenso m\'as pronunciado, la contribuci\'on de los niveles de Landau excitados ($l \geq 1$) es:
    
    \begin{align}
    \Omega_{F}^{l\neq 0} &\approx - \frac{eB}{2\pi} \sum_{l=1}^{l_{max}} \alpha_l \Bigg[ \int_0^{p_F^{(l)}} dp_3 (E_l - \mu_F) \nonumber + \frac{\pi^2 T^2}{6} \left( \frac{dN_l}{dE_l} \Bigg|_{E_l=\mu} \right) \Bigg],
    \end{align}
    
\noindent donde $p_F^{(l)} = \sqrt{\mu^2 - m_F^2 - 2eBl}$ es el momento de Fermi para el nivel $l$, y $n_l$ es la densidad de estados y $l_{max}=I[\frac{mu^2-m^2}{2eB}]$ es el m\'aximo n\'umero de niveles de Landau ocupados para un campo magn\'eticoy potencial qu\'imico fijos. $I[z]$ denota la parte entera de $z$.
    
    Para campos magn\'eticos intermedios ($m_F^2 \lesssim 2eBl \lesssim \mu^2$), la contribuci\'on puede expresarse como:
    
    \begin{equation}
    \Omega_{F}^{l\neq 0} = - \frac{(eB)^{3/2} T^{5/2}}{(2\pi)^{3/2}} \sum_{k=1}^\infty \frac{(-1)^{k+1}}{k^{5/2}} \cosh\left( \frac{k\mu}{T} \right) e^{-k m_F^2/(2eB)} \text{Li}_{1/2}(e^{-k eBl/T}).
    \end{equation}

Para fermiones neutros podemos proceder de manera an\'aloga considerando el espectro de energ\'a de fermiones neutros. Estas part\'iculas interactuar\'an con el campo magn\'etico a trav\'es del momento magn\'etico. 
La forma expl\'icita del potencial termodin\'amico toma la forma

    \begin{equation}
        \Omega_{n, st}^{LT} = \frac{(m_F T)^{3/2}}{\sqrt{2}\pi^2} [Li(z_{+}) + Li(z_{-})].
    \end{equation}
   
\section{Ecuaciones de Estado anisotr\'opicas: Gas mixto \texorpdfstring{$npe + \pi^-$}{npe + pi-}}

    Esta secci\'on tiene como objetivo obtener las ecuaciones de estado a $T = 0$ de un gas \textit{npe}, con piones cargados incluidos.

    Para calcular las ecuaciones de estado del sistema   partiremos del tensor de energ\'ia-momento. La parte diagonal del tensor,  es el tensor de esfuerzo, cuya parte espacial contiene a las presiones y la componente temporal a la densidad de energ\'ia interna $E$. A partir del potencial termodin\'amico $T^i_j$ puede calcularse como: 
    
    \begin{equation}\label{emtensor}
    T^i_j=\frac{\partial\Omega}{\partial a_{i,\lambda}}a_{j,\lambda}-\Omega\delta_{j}^i,\quad\quad T_4^4=-E,
    \end{equation}
    
    \noindent donde $a_{i}$ denota los campos presentes (fermiones, bosones, electromagn\'etico, etc.) \cite{PerezRojas:2006dq}.
    Para un potencial termodin\'amico que depende de un campo externo, la Ec.(\ref{emtensor}) conduce a t\'erminos de presi\'on con la forma:
    
    \begin{equation}
    \textit{T}^i_j=-\Omega-F_k^i\left( \frac{\partial\Omega}{\partial F_k^j}\right),\quad  i=j.
    \end{equation}
    
    Teniendo en cuenta que el campo magn\'etico est\'a dirigido en la direcci\'on $\textbf{e}_3$,  la anisotrop\'ia de las presiones  se hace expl\'icita\cite{PerezRojas:2006dq}
    
    \begin{subequations}\label{pressures}
    	\begin{align}
    P_{\parallel}&=\textit{T}^3_3=-\Omega = -\Omega_{st} -\Omega_{vac},\\
    P_{\perp}&=\textit{T}_1^1=\textit{T}_2^2=-\Omega-B \mathcal M = P_{\parallel}-B \mathcal M. 
    \end{align}
    \end{subequations}
    
La presencia de un campo magn\'etico rompe la simetría rotacional, haciendo que los componentes espaciales del valor estadístico promedio del tensor energía-momento se vuelvan anisotr\'opicos\cite{DaryelModelingCO}. En consecuencia, la ecuaci\'on de estado (EoS) se vuelve anisotr\'opica, con la presi\'on dividi\'endose en componentes \textit{paralela} ($\hat{P}_\parallel$) y \textit{perpendicular} ($\hat{P}_\perp$) respecto a la direcci\'on de las líneas del campo magn\'etico. Estas presiones son:
	\begin{align}
		\hat{P}_\parallel &= -\hat{\Omega}, \label{eq:p_parallel} \\
		\hat{P}_\perp &= -\hat{\Omega} - \hat{\mathcal{M}} b, \label{eq:p_perp}
	\end{align}
\noindent donde $\hat{\Omega}$ y $\hat{\mathcal{M}}$ son el potencial termodin\'amico total y la magnetizaci\'on (incluyendo las contribuciones del vacío y estadística).
    
La presencia del campo magn\'etico en el sistema produce un colapso magn\'etico transversal  dado por la condici\'on 
    \begin{equation} \label{colapsop}
    P_{\perp}=-\Omega-{\mathcal M} B\leq 0,
    \end{equation}
 En una estrella este colapso magn\'etico  actuar\'ia como un desencadenante de la expulsi\'on de la materia a trav\'es del colapso magn\'etico transversal para uno o varios de los gases que componen la estrella. 
Como nuestro inter\'es y motivaci\'on es astrof\'isica, en lo que sigue estudiaremos  una mezcla de gases ideales de electrones, protones, neutrones y piones que  interact\'uan con el campo magn\'etico. Para los piones  haremos la excepci\'on de considerar  que adem\'as interact\'uan entre ellos.

\subsection{\texorpdfstring{Gas $npe+\pi^-$ a $T=0$, $\mu \ne 0$}{Gas npe+pi- a T=0, mu ≠ 0}}

Para un gas de fermiones cargado, las EdE en el límite $T = 0$ son \cite{GonzalezFelipe:2021IJMPD}:    
\begin{subequations}\label{EoSCF}
\begin{align}
E^{e,p} &= \frac{m_{e,p}^2}{4\pi^2}\frac{B}{B^{e,p}_c} \sum_{l=0}^{l_{\text{max}}} g_{l} \left( \mu^{e,p}\,p^{e,p}_F + {\varepsilon_{l}}^2 \ln\frac{ {\mu^{e,p}}+ {p^{e,p}_F}}{\varepsilon_{l}}\right), \label{EoSCF:a} \\
P^{e,p}_{\parallel} &= \frac{m_{e,p}^2}{4\pi^2}\frac{B}{B^{e,p}_c} \sum_{l=0}^{l_{\text{max}}} g_{l} \left[ \mu^{e,p} \,p^{e,p}_F -{\varepsilon_{l}}^2 \ln\left(\frac{\mu^{e,p}+ p^{e,p}_F}{\varepsilon_{l}}\right)\right], \label{EoSCF:b} \\
P^{e,p}_{\perp} &= \frac{m_{e,p}^4}{2\pi^2} \left(\frac{B}{B^{e,p}_c}\right)^2 \sum_{l=0}^{l_{\text{max}}} g_l\, l \ln\left (\frac{\mu^{e,p}+p^{e,p}_F}{\varepsilon_l}\right), \label{EoSCF:c} \\
\mathcal{M}^{e,p} &= \frac{m_{e,p}^2}{4\pi^2 B_c^{e,p}} \sum_{l=0}^{l_{\text{max}}} g_{l} \left[ \mu^{e,p} \varepsilon_{l} - \varepsilon_{l} \ln \left(\frac{\mu^{e,p} + \varepsilon_{l}}{\varepsilon_{l}}\right) \right], \label{EoSCF:d} \\
N^{e,p} &= \frac{m_{e,p}^2}{4\pi^2} \frac{B}{B^{e,p}_c} \sum_{l=0}^{l_{\text{max}}} 2 g_{l} p^{e,p}_F. \label{EoSCF:e}
\end{align}
\end{subequations}
    
    \noindent donde $B^{e,p}_c = m_{e,p}^2/e^{e,p}$.
    
    Por su parte las EdE del gas magnetizado de neutrones se reducen a $T=0$ a las siguientes expresiones \begin{subequations}\label{EoSNF}
\begin{align}
E^{n} &= - P^{n}_{\parallel} + \mu^{n} N^{n}, \label{EoSNF:a} \\
P^{n}_{\parallel} &=  \frac{m_{n}^4}{2\pi^2} \sum_{\eta=1,-1} \left\{ \frac{\mu^{n} f^3}{12 m_n} + \frac{(1+\eta b^n)(5 \eta b^{n} -3)\mu^n f}{24 m_{n}} + \frac{(1+\eta b^{n})^3 L}{24} - \frac{\eta b^{n} (\mu^{n})^3 s}{6 m_{n}^3} \right\}, \label{EoSNF:b} \\
P^{n}_{\perp} &=  P^{n}_{\parallel} - \mathcal{M}^{n} B, \label{EoSNF:c} \\
\mathcal{M}^{n} &=   \frac{m_{n}^3 \kappa_n}{2\pi^2} \sum_{\eta=1,-1} \eta \left\{ \frac{(1-2 \eta b^{n})}{6} f + \frac{(1 + \eta b^{n})^2(1 - \eta b^{n}/2)}{3} L - \frac{(\mu^{n})^3}{6 m_{n}^2} s \right\}, \label{EoSNF:d} \\
N^{n} &=  \frac{m_{n}^3}{2\pi^2} \sum_{\eta=1,-1} \left\{ \frac{f^3}{3} + \frac{\eta b^{n}(1 + \eta b^{n})}{2}f - \frac{\eta b^{n} (\mu^{n})^2}{2 m_{n}^2}s \right\}. \label{EoSNF:e}
\end{align}
\end{subequations}
    
    \noindent donde $b^{n}=B/B^{n}_c$ con $B^{n}_c=m_{n}/\kappa_{n}$, siendo $m_n$ y $\kappa_n$, la masa y el momento magn\'etico del neutr\'on. Las funciones $f$, $L$ y $s$ se definen como:
    \begin{eqnarray}\label{EoSNFaux}
    f &=& \frac{(\mu^n)^2-(\varepsilon^n(\eta))^2}{m_n},\\
    L &=& \frac{1}{1+\eta b^n} \ln \left(\frac{\mu^n + \sqrt{(\mu^n)^2 - \varepsilon^n(\eta)^2}}{m_n}\right),\\
    s &=& \frac{\pi}{2}-\frac{m_n}{\mu^n} \arcsin(1+\eta b^n),
    \end{eqnarray}
con $\epsilon^n = m_n - \eta \kappa B$.
    Los piones cargados, a $T = 0$ est\'an condensados y por lo tanto, la contribuci\'on del  potencial termodin\'amico esta\'istico a las presiones paralela y perpendicular es nula (ver ecuaci\'on \eqref{Omegab}). El vac\'io por su parte contribuye a la presiones paralela y perpendicular.  Ec.~(\ref{Pot_Vacio}a). 
    Sin embargo la presi\'on perpendicular tiene la contribuci\'on de la magnetizaci\'on que es no nula a $T=0$ \eqref{magneLLL}. La contribuci\'on de la densidad de energ\'ia es tambi\'en no nula como se aprecia de la Ec. \eqref{EnerLLL}.
    
    Para estrellas de condensados de Bose Einstein,  se toma en cuenta  una interacci\'on simple, de tipo contacto $U_0\delta(r-r^{\prime})$ y r y $r^{\prime}$ denotan las posiciones de las part\'iculas interactuantes. \cite{QuinteroAngulo2022}-\cite{Adriel}-\cite{Chavanis2012}. El par\'ametro   $U_0 = \frac{2\pi a}{m_{\pi}}$, con $m_{\pi}$ la masa del pion y $a$ es la longitud de dispersi\'on.    Esta interacci\'on se añade a las ecuaciones de estado, a la energ\'ia y a las presiones. Las ecuaciones de estado quedan entonces

    \begin{subequations}\label{EdEpi}
    \begin{align}
    E^{\pi} &= -\frac{e B \, \left(m^2 + eB\right)^{3/4}}{2\sqrt{2}\pi m^{3/2}} + \frac{1}{2}U_0N^2, \label{EdEpi:a} \\
    P_\parallel^\pi(B) &= -\frac{(eB)^2}{16\pi^2} \Bigg[ 
    -\frac{\gamma_E - \ln\left(\frac{Be}{m^2}\right)}{6} + \frac{m^4}{4B^2 e^2} \left(-3 + 2\gamma_E - \ln\left(\frac{B^2 e^2}{m^4}\right)\right) \nonumber \\
    &+ 4(-1 + \gamma_E + \ln 2) \, \zeta\left(-1, \frac{Be + m^2}{2Be}\right) - 4 \, \zeta^{(1,0)}\left(-1, \frac{Be + m^2}{2Be}\right) \Bigg] + \frac{1}{2}U_0N^2, \label{EdEpi:b}\\
    P_\perp^\pi &= P_\parallel^\pi - B M^\pi, \label{EdEpi:c} \\
    \mathcal{M}_{\pi} &= -\frac{eN^\pi}{2\sqrt{m^2 + eB}}, \label{EdEpi:d}
    \end{align}
    \end{subequations}
    
   La densidad  de bosones $N^\pi$ se tomar\'a de modo tal que satisfaga la condici\'on de neutralidad de carga total del gas.
    
    \subsection{Condiciones de equilibrio estelar}

    Derivaremos ahora las condiciones de equilibrio estelar que deben cumplirse si la materia que forma el interior del objeto compacto es sistema \textit{npe} con piones cargados. 
   
    
    Si ignoramos por un momento las fuertes interacciones de los piones con la materia que los rodea, vemos que es posible que aparezcan $\pi^-$ en materia rica en neutrones mediante la reacci\'on $n \rightarrow p + \pi^-$ una vez que el potencial químico del neutr\'on, $\mu_n$, excede el potencial químico del prot\'on, $\mu_p$, al menos en la masa en reposo del $\pi^-$, $m_\pi = 139.6\,\text{MeV}$. En la materia de las estrellas de neutrones, que est\'a en equilibrio beta, la diferencia de potenciales químicos $\mu_n - \mu_p$, es igual al potencial químico del electr\'on $\mu_e$, alcanza los $110\,\text{MeV}$ a la densidad de la materia nuclear$^1$, y uno podría esperar la aparici\'on de $\pi^-$ a densidades ligeramente mayores \cite{Pajkos2024}.
    
    Sin embargo, no se pueden ignorar las interacciones del pi\'on con su entorno. La fuerte repulsi\'on en onda s entre neutrones y piones tiende a aumentar la masa efectiva del pi\'on y, por lo tanto, eleva la densidad umbral para la aparici\'on de piones \cite{Pajkos2024}. A bajas energías del $\pi^-$, este aumento en la energía debido a las interacciones en onda s entre piones y nucleones puede estimarse f\'acilmente en t\'erminos de las longitudes de dispersi\'on medidas.
    
    El equilibrio bajo la reacci\'on $n \leftrightarrow p + \pi^-$ implica que $\mu_n = \mu_p + \mu_\pi$, donde $\mu_n$ y $\mu_p$ son los potenciales químicos del neutr\'on y el prot\'on, por tanto, tenemos:
    \begin{align}\label{BetaEq}
    \mu_{n} &= \mu_{p} + \mu_{e},\\
    \mu_{\pi} &= \mu_{e}.
    \end{align}
     
    Adem\'as del equilibrio de interacci\'on d\'ebil, tambi\'en debemos considerar la neutralidad de carga. Por tanto, las fracciones de part\'iculas con carga positiva y negativa deben igualarse:
    
    \begin{equation}\label{ChargeNeutrality}
    N^{\pi} = N^{p} - N^{e},
    \end{equation}

    Los potenciales químicos del prot\'on y el electr\'on se determinan unívocamente a partir de estas relaciones. Teniendo en cuenta adem\'as la conservaci\'on del n\'umero bari\'onico $n_B$:
    \begin{equation}\label{barion}
    n_{B} = N^{n} + N^{p}.
    \end{equation}

    \begin{figure}[!ht]
	\centering
	\includegraphics[width=0.46\linewidth]{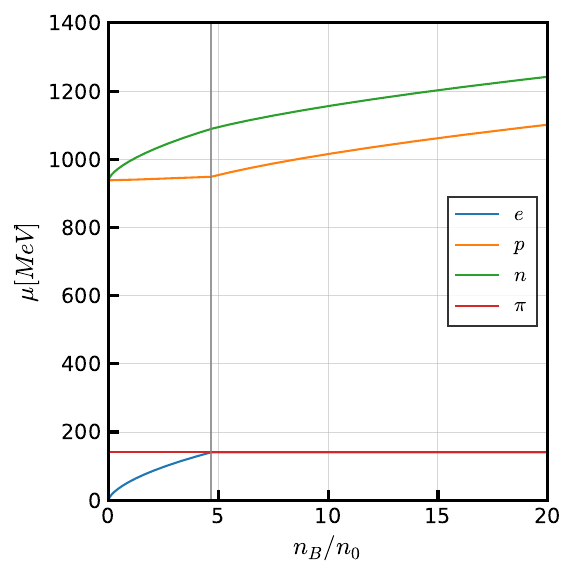}
	\includegraphics[width=0.46\linewidth]{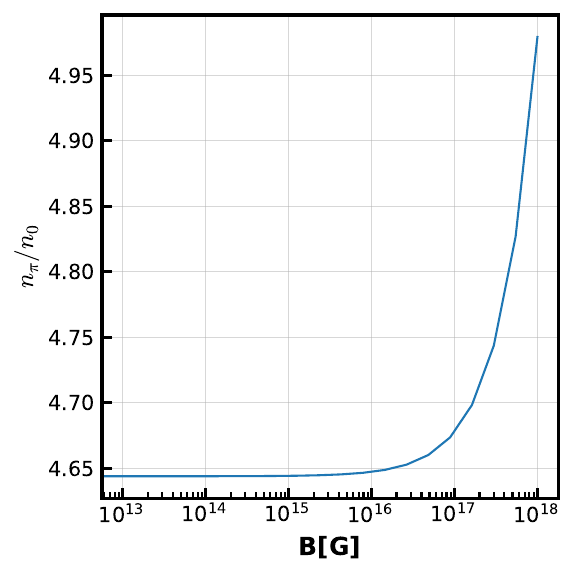}
	\caption{EL panel izquierdo muestra el comportamiento de los potenciales químicos resultantes de resolver las ecuaciones de equilibrio para $B=10^{17}$ G contra la densidad bari\'onica del sistema. El panel derecho muestra la densidad a la cual aparecen los piones para un campo magn\'etico dado.}\label{densidad}
    \end{figure}
    Las Ecs (\ref{BetaEq}, \ref{ChargeNeutrality} y \ref{barion}) constituyen nuestras condiciones de equilibrio estelar. La Figura [\ref{densidad}] muestra c\'omo estas ecuaciones de equilibrio condicionan las cantidades de las part\'iculas que conforman el gas. En el panel izquierdo, se ve el comportamiento de los potenciales qu\'imicos de cada especie como funci\'on de la densidad bari\'onica adimensionalizada con respecto a la densidad de saturaci\'on nuclear $n_0$, para un campo magn\'etico $B = 10^{17}G$, resaltando el valor de esta para el cual comienzan a aparecer los piones en el gas. El panel derecho muestra la densidad a la que comienzan a aparecer los piones para un $B$ dado.

    \subsection{Ecuaciones de Estado del gas mixto}
    
    Las ecuaciones de estado de la materia al interior del OC, considerando las condiciones de equilibro estelar puedes escribirse de manera resumida como cantidades totales de la mezcla, en particular las presiones totales, $P^T_{\parallel}$ y $P_{\perp}^T$ y la densidad de energía total $E^T$, se calculan como la suma de las contribuciones de cada gas, dadas por las Ecs. ~\eqref{EdEpi}, \eqref{EoSCF} y \eqref{EoSNF}. Adem\'as, las EdE deben contener las contribuciones energ\'eticas de los campos presentes en el sistema, excepto la correspondiente al campo gravitacional \cite{Schmitt2010}. Por ello, para completar las EdE es necesario a\~nadir la contribuci\'on del campo magn\'etico \cite{NSObserations}. Desde el punto de vista matem\'atico, para ello basta a\~nadir la llamada contribuci\'on de Maxwell, $B^2/8 \pi$, a $E$ y $P_{\perp}$, y sustraerla de $P_{\parallel}$, con lo que obtenemos las siguientes EdE para el gas mixto:
\begin{align}
    E^T &= \sum_{i = e, p, n, \pi} E^i + \frac{B^2}{8\pi}, \\
    P^T_{\parallel} &= \sum_{i = e, p, n, \pi} P^i_{\parallel} - \frac{B^2}{8\pi},\\
    P^T_{\perp} &= \sum_{i = e, p, n, \pi} P^i_{\perp} + \frac{B^2}{8\pi}.
\end{align}\label{edenpepi}
    \chapter{Ecuaciones de estructura anisotr\'opicas}\label{cap3}

En este cap\'itulo, se discute el problema de determinar la estructura (masa, forma y dimensiones) de objetos compactos anisotr\'opicos. En una primera parte se presenta la manera cl\'asica de atacar este problema en el caso de objetos compactos esf\'ericos a trav\'es de las ecuaciones de Tolman-Oppenheimer-Volkoff\cite{Tolman, OppenheimerVolkoff},. Posteriormente, se describe el proceso para la obtenci\'on y uso de ecuaciones de estructura para objetos esferiodales, partiendo de una m\'etrica axisim\'etrica que es una generalizaci\'on de la m\'etrica de Schwarzschild.

\section{Equilibrio hidrost\'atico estelar}

Debido a las altas densidades que se dan en los objetos compactos y en particular, en las Estrellas de Neutrones (de bosones en nuestro caso), las correcciones de la relatividad general a las ecuaciones de Newton son importantes para la descripci\'on del equilibrio hidrost\'atico de estos objetos \cite{Shapiro1983,camenzind2007compact}. Cuando en el marco de la Teor\'ia de la Relatividad General se habla de modelos de estrellas, a lo que se hace referencia es a una regi\'on interior que es una soluci\'on de las ecuaciones de Eisntein con fuente de materia ($T_{\mu\nu} \neq 0$), y una regi\'on exterior cuya m\'etrica es una soluci\'on asint\'oticamente plana de estas mismas ecuaciones en el vac\'io ($T_{\mu \nu} = 0$) \cite{camenzind2007compact}. Estas dos piezas deben ser cuidadosamente empatadas en la superficie de la estrella.

El problema de encontrar soluciones exactas de las ecuaciones de Einstein para las fuentes de materia especificadas por unas EdE no es trivial, incluso en el caso m\'as sencillo de materia en simetr\'ia esf\'erica \cite{Malafarina2005}. Una manera de abordarlo consiste en proponer una m\'etrica interior a partir de ciertas consideraciones generales y luego obtener $T_{\mu \nu}$ a trav\'es de las ecuaciones de Einstein \cite{Malafarina2005}. 

Los intentos de dar soluci\'on a este problema de manera inversa, es decir, de determinar la m\'etrica a partir de las EdE, han sido igualmente numerosos. Si las EdE son isotr\'opicas esta estrategia conduce a las ecuaciones de Tolman-Oppenheimer-Volkoff que permiten la obtenci\'on de observables macrosc\'opicos (masas y radios) \cite{OppenheimerVolkoff,Tolman,camenzind2007compact}.
En el caso de ecuaciones de estado anisotr\'opicas que conducir\'ian a objetos compactos no esf\'ericos, los estudios se centran m\'as en la obtenci\'on de las propiedades matem\'aticas de las soluciones de las ecuaciones de Einstein dada una simetr\'ia, que en la obtenci\'on de observables macrosc\'opicos derivados de una EdE espec\'ifica \cite{OppenheimerVolkoff,Tolman, Malafarina2005,esposito}.

\section{Ecuaciones de Tolman-Oppenheimer-Volkoff}

La fuente de la curvatura del espacio-tiempo es la materia, cuya descripci\'on física viene dada por el tensor energía-momento. La interacci\'on entre el espacio-tiempo y la materia es a trav\'es de las ecuaciones de Einstein

\begin{equation}\label{TGR}
G_{\nu}^{\mu} = R_{\nu}^{\mu} - \frac{1}{2} g_{\nu}^{\mu} R = -\kappa T_{\nu}^{\mu},
\end{equation}

donde $G_{\nu}^{\mu}$ es el tensor de Einstein que describe la curvatura del espacio-tiempo, $T_{\nu}^{\mu}$ es el tensor energía-momento, $R_{\nu}^{\mu}$ es el tensor de Ricci, $R$ el escalar de Ricci y $\kappa = 8\pi G_{N}$ es una constante de proporcionalidad, siendo $G_{N} = 6,\!711 \times 10^{-45} \, \text{MeV}^{-2}$ la constante de gravitaci\'on universal. Adem\'as:

\begin{equation}
R_{\mu\nu} = \Gamma^{\alpha}_{\mu\nu,\alpha} - \Gamma^{\alpha}_{\mu\alpha,\nu} + \Gamma^{\alpha}_{\mu\nu}\Gamma^{\beta}_{\alpha\beta} - \Gamma^{\beta}_{\mu\alpha}\Gamma^{\alpha}_{\nu\beta},    
\end{equation}

donde
\begin{equation}
\Gamma^{\alpha}_{\mu\nu} = \frac{g^{\alpha\beta}}{2} \left( \frac{\partial g_{\beta\mu}}{\partial x^{\nu}} + \frac{\partial g_{\nu\beta}}{\partial x^{\mu}} - \frac{\partial g_{\mu\nu}}{\partial x^{\beta}} \right).
\end{equation}

son los símbolos de Christoffel y $g_{\mu\nu}$ el tensor m\'etrico

Como todos los tensores que aparecen en las Ecs.~(\ref{TGR}) son sim\'etricos, ellas  tienen diez componentes independientes que, dada la libertad de elecci\'on de las cuatro coordenadas del espacio-tiempo, se reducen a seis. De manera que las ecuaciones de Einstein son un sistema de seis ecuaciones diferenciales en derivadas parciales no lineales.
 
En el caso de un objeto compacto est\'atico cuya materia se comporta como un fluido perfecto ($T_{\mu\nu}=diag(-E,P,P,P)$), el alto nivel de simetr\'ia determina que todos los elementos no diagonales de la m\'etrica del espacio-tiempo que ellos generan sean nulos, y que los diagonales dependan \'unicamente de la distancia al origen de coordenadas. En consecuencia, la m\'etrica del espacio-tiempo generado por una estrella est\'atica y esf\'erica puede escribirse como \cite{camenzind2007compact}:

\begin{equation}\label{Sch0}
ds^2 = - e^{2 \Phi(r)} dt^2 + e^{2 \Lambda(r)} dr^2
+ r^{2} d\theta^2
+ r^2\sin^2\theta d\phi^2,
\end{equation}

\noindent donde $r,\theta,\phi$ son las coordenadas esf\'ericas usuales. Las funciones $\Phi(r)$ y $\Lambda(r)$ se determinan un\'ivocamente a partir de la distribuci\'on de energ\'ia en el interior de la estrella.

Con el uso de la Ec.~(\ref{Sch0}) los s\'imbolos de Christoffel $\Gamma^{\alpha}_{\mu\nu}$, el tensor y el escalar de Ricci, $R_{\mu\nu}$ y $R=R^{\mu}_{\,\,\,\,\mu}$ respectivamente, y el tensor de Einstein $G_{\mu \nu}$, pueden calcularse como funci\'on de $\Phi(r)$, $\Lambda(r)$ y $r$. Una vez que $G_{\mu\nu}$ es puesto en funci\'on de la m\'etrica, sus componentes se sustituyen en las Ecs.(\ref{TGR}), de conjunto con el tensor energ\'ia-momento de la materia en cuesti\'on, que en este caso es $T_{\mu\nu}=diag(-E,P,P,P)$. Luego de varias transformaciones algebraicas, las ecuaciones de Einstein se reducen a las cuatro ecuaciones diferenciales \cite{camenzind2007compact}:

\begin{subequations}\label{TOV}
	\begin{align}
\frac{dm(r)}{dr} &= 4 \pi r^2 E(r),\label{TOV1}\\	
\frac{dP(r)}{dr}&=-G\frac{(E(r)+P(r))\left( 4 \pi r^{3} P(r) + m(r)\right)}{ r^{2}\left(1-\frac{2Gm(r)}{r}\right)}, \label{TOV2}\\
e^{-2\Lambda(r)} &= \left(1-\frac{2Gm(r)}{r}\right), \label{TOV3}\\
\frac{d\Phi(r)}{dr} &=\frac{G}{\left(1-\frac{2Gm(r)}{r} \right)} \left(\frac{m(r)}{r^2} + 4 \pi r P(r)\right) ,\label{TOV$}
\end{align}
\end{subequations}

\noindent que describen la dependencia de la masa $m(r)$, la presi\'on $P(r)$, y las dos funciones m\'etricas $\Phi(r)$ y $\Lambda(r)$ con el radio interno de la estrella $r$. Las cuatro Ecs.~(\ref{TOV}) determinan completamente la estructura del objeto compacto y son conocidas como ecuaciones de Tolman-Oppenheimer-Volkoff (TOV), aunque por lo general, cuando se habla de las TOV se hace referencia solamente a las dos primeras ya que en la pr\'actica, para obtener la masa y el radio de la estrella basta con resolver estas dos ecuaciones.

Para resolver las Ecs.~(\ref{TOV1}) y (\ref{TOV2}) se parte de la condiciones en el centro de la estrella $m(r=0)=0$ y $P(r=0)=P_0$, donde $P_0$ se toma de la ecuaci\'on de estado. Posteriormente se integra en la variable $r$ hasta que la presi\'on se hace igual a cero. La condici\'on $P(R)=0$ define el radio de la estrella, que permite a su vez determinar su masa total $m=m(R)$. Al resolver las Ecs.~(\ref{TOV1}) y (\ref{TOV2}) en el rango de densidades admitido por las EdE, se obtiene una familia \'unica de estrellas, una curva masa-radio, parametrizada por la presi\'on y la densidad de energ\'ia centrales $(E_0,P_0)$. Cada punto $(M(E_0),R(E_0))$ de la familia obtenida respresenta  una estrella de masa $M$ y radio $R$ en equilibrio hidrodin\'amico. En una secuencia estelar, solamente las ramas en las que $dm/dE_0 > 0$ son estables ante perturbaciones radiales \cite{camenzind2007compact,Shapiro1983}.

Aunque las ecuaciones TOV han sido cruciales en el estudio y compresi\'on de la f\'isica de los objetos compactos, el suponer la estrella esf\'erica implica que las ecuaciones (\ref{TOV}) no pueden, en rigor, ser utilizadas para el estudio de objetos compactos magnetizados, pues ellas no pueden tener en cuenta las deformaciones producidas porla separaci\'on de las presiones en paralela y perpendicular. 

\section{Ecuaciones de estructura anisotr\'opicas: M\'etrica \texorpdfstring{$\gamma$}{gamma}}

En un intento de describir adecuadamente la estructura macrosc\'opica de los OCs magnetizados, derivamos, un conjunto de ecuaciones de estructura tipo TOV a partir de una m\'etrica axialmente sim\'etrica en coordenadas esf\'ericas ($t, r, \theta, \phi$) \cite{Zubairi2017,ZubairiRomero_2015, WhiteGamma}, la m\'etrica $\gamma$, cuya forma en coordenadas esf\'ericas $(t,r,\theta,\phi)$ es:

\begin{equation}\label{gammatotal}
ds^2 = - \Delta^\gamma dt^2 + \Delta^{\gamma^2-\gamma-1} \Sigma^{1-\gamma^2} dr^2
+ r^{2} \Delta^{1-\gamma} \Sigma^{1-\gamma^2} d\theta^2
+ r^2\sin^2\theta \Delta^{\gamma^2-\gamma} d\phi^2,
\end{equation}

\noindent con:

\begin{subequations}\label{deltasigma}
	\begin{align}
	\Delta& = \left ( 1 - \frac{2 G m}{r}\right),\\
	\Sigma& = \left ( 1 - \frac{2 G m}{r} + \frac{G^2 m^2}{r^2} \sin^2 \theta \right)
	\end{align}
\end{subequations}

La m\'etrica $\gamma$ es una familia de soluciones est\'aticas, axisim\'etricas y asint\'oticamente planas de las ecuaciones de Einstein \cite{Malafarina2005,esposito}. Ella depende de dos par\'ametros, $m$ y $\gamma$. El sentido f\'isico de los par\'ametros $m$ y $\gamma$ pueden ser investigado a partir de los momentos monopolar $M$ y cuadrupolar $Q$ de la m\'etrica $\gamma$ \cite{esposito}:

\begin{align}
    M &= \gamma m, \\
    Q &= \frac{1}{3} m^3 \gamma (1-\gamma^2),
\end{align}\label{masspole}

\noindent De las Ecs.~(\ref{masspole}) se sigue que el par\'ametro $m$ est\'a relacionado con la masa total del objeto, mientras que el par\'ametro $\gamma$ tiene que ver con su forma \cite{Malafarina2005}. N\'otese que para $\gamma = 0$, la Ec.(\ref{gammatotal}) se reduce al espacio-tiempo plano de Minkowski, con $M=Q=0$. Por otra parte, si $\gamma = 1$, la Ec.(\ref{gammatotal}) se reduce a la m\'etrica de Schwarzschild y la simetr\'ia esf\'erica se recupera. Esto significa que el objeto en cuesti\'on no est\'a deformado pues ahora nuevamente $Q=0$ como corresponde a una fuente de masa esf\'erica.

Como la m\'etrica $\gamma$ puede ser llevada de manera continua hacia la de Schwarzschild, un primer paso razonable hacia la construcci\'on de ecuaciones de estructura para objetos no esf\'ericos, es partir de la Ec.~(\ref{gammatotal}) en el l\'imite $\gamma \cong 1$, lo que es equivalente a considerar objetos poco deformados \cite{Malafarina2005}. Para $\gamma \cong 1$, las Ecs.~(\ref{gammatotal})-(\ref{deltasigma}) se reducen a:

\begin{equation}\label{gammauno}
ds^2 = - \left[1-2G m(r)/r\right]^{\gamma}dt^2 +  \left[1-2G m(r)/r\right]^{-\gamma}dr^2
+ r^2\sin^2\theta d\phi^2 + r^{2}d\theta^2.
\end{equation}

Siguiendo el procedimiento explicado en la secci\'on anterior, las ecuaciones de Eisntein se resuelven utilizando la m\'etrica Ec.~(\ref{gammauno}) y suponiendo el tensor de energ\'ia-momento isotr\'opico, se obtiene la siguiente ecuaci\'on de estructura para el objeto compacto deformado \cite{ZubairiRomero_2015}:
\begin{equation}\label{gTOV0}
	\frac{dP}{dr}=- \frac{(E+P)\left[\frac{r}{2}+4\pi r^{3} G  P-\frac{r}{2}\left(1-\frac{2Gm}{r}\right)^{\gamma}\right]}{ r^{2}\left(1-\frac{2Gm}{r}\right)^{\gamma}}.
\end{equation}
En la Ec.~(\ref{gTOV0}), la deformaci\'on entra solo a trav\'es del par\'ametro $\gamma$, y aunque ella se reduce a la Ec.~(\ref{TOV}) cuando $\gamma = 1$. Al obtenerse suponiendo el tensor de energ\'ia-momento isotr\'opico, la Ec.~(\ref{gTOV0}) no permite tampoco, en principio, el uso de presiones anisotr\'opicas.

A fin de obtener un conjunto de ecuaciones de estructura que permita tener en cuenta el car\'acter anisotr\'opico de las ecuaciones de estado, la Ec.~(\ref{gTOV0}) debe ser complementada con la informaci\'on que las EdE aportan sobre el objeto compacto. Siendo el empuje de la presi\'on hacia afuera lo que impide el colapso gravitatorio, que la presi\'on ejercida por el gas sea diferente en las direcciones polar (paralela) y ecualtorial (perpendicular) implica que las dimensiones de la estrella a lo largo de estas direcciones son distintas. A fin de incluir las dos presiones en la descripci\'on de la estructura del objeto, se propone en primer lugar, que el objeto compacto es esferoidal. En segundo lugar, se propone \cite{ZubairiRomero_2015} parametrizar la coordenada que describe la distancia en el eje polar $z$ en t\'erminos de $\gamma$ y la coordenada ecuatorial:

\begin{equation}\label{gammazr}
z = \gamma r,
\end{equation}

\noindent aprovechando adem\'as el hecho de que $\gamma$ est\'a relacionada con la deformaci\'on del objeto. Con ayuda de la Ec.~(\ref{gammazr}), la masa contenida en un esferoide de radio ecuatorial $r$ y radio polar $z=\gamma r$  puede ser calculada como \cite{ZubairiRomero_2015}:

\begin{equation}\label{masa}
\frac{dm}{dr} = 4 \pi \gamma r^2 E,
\end{equation}

\noindent y la Ec.~(\ref{gTOV0}) puede reescribirse en funci\'on del radio polar $z$ \cite{Zubairi2017}:

\begin{equation}\label{gTOVz}
\frac{dP}{dz}=-\frac{(E+P)\left[\frac{z}{2 \gamma}+4 \pi G  \left(\frac{z}{\gamma}\right)^{3}P-\frac{z}{2 \gamma}\left(1-\frac{2 G m \gamma}{z}\right)^{\gamma}\right]}{ \left(\frac{z}{\gamma}\right)^{2}\left(1-\frac{2 G m\gamma}{z}\right)^{\gamma}}.
\end{equation}

Llegados a este punto, la anisotrop\'ia en las presiones puede introducirse en las ecuaciones de estructura \cite{Zubairi2017}. Para ello se supone que la presi\'on paralela est\'a asociada \'unicamente con la coordenada $z$, mientras que la perpendicular lo estar\'ia solo con $r$. Haciendo esto, se llega al sistema de ecuaciones diferenciales:

\begin{subequations}\label{gTOVpartial}
	\begin{align}
\frac{dm}{dr} &= 4 \pi \gamma r^2 E,\label{gTOVpartial1}  \\
\frac{dP_{\parallel}}{dz}&=-\frac{(E+P_{\parallel})\left[\frac{z}{2 \gamma}+4\pi G \left(\frac{z}{\gamma}\right)^{3}P_{\parallel}-\frac{z}{2 \gamma}\left(1-\frac{2 G m \gamma}{z}\right)^{\gamma}\right]}{ \left(\frac{z}{\gamma}\right)^{2}\left(1-\frac{2 G m\gamma}{z}\right)^{\gamma}}, \label{gTOVpartial2}\\
	\frac{dP_{\perp}}{dr}&=-\frac{(E+P_{\perp})\left[\frac{r}{2}+4 \pi G r^{3}P_{\perp}-\frac{r}{2}\left(1-\frac{2 G m}{r}\right)^{\gamma}\right]}{ r^{2}\left(1-\frac{2 G m}{r}\right)^{\gamma}},\label{gTOVpartial3}	\end{align}
\end{subequations}

\noindent que describen la variaci\'on de la masa y las presiones con las coordenadas espaciales $r$ y $z$ para un objeto compacto esfeoridal. N\'otese que las Ecs.~(\ref{gTOVpartial2}) y (\ref{gTOVpartial3}) est\'an acolpladas entre s\'i a trav\'es de su dependencia con la densidad de energ\'ia $E$, y la masa $m(r)$. Para una descripci\'on detallada del proceso de integraci\'on de las Ecs (\ref{gTOVpartial}), ver Ap\'endice \ref{appB}.


    \chapter{Resultados num\'ericos y discusi\'on}\label{cap4}

En este cap\'itulo se estudian los efectos del campo magn\'etico en la estructura, masa, radio y deformaci\'on, de Estrellas de condensado de Bose-Einstein. Para ello se combinan las ecuaciones de estado obtenidas en el Cap\'itulo \ref{cap3} con las ecuaciones de estructura derivadas en el Cap\'itulo \ref{cap4}. Como en el Cap\'itulo \ref{cap3}, el estudio se har\'a suponiendo que el campo magn\'etico es constante en el interior de la estrella.

\section{Relaci\'on masa-radio y masa-energ\'ia}

El objetivo \'ultimo de la modelaci\'on de cualquier objeto compacto es la obtenci\'on de sus observables, pues ellos no solo permiten validar el modelo en cuesti\'on a trav\'es de compararlos con datos observacionales, sino que son cruciales para la interpretaci\'on de estos \'ultimos. Los observables derivados de un modelo espec\'ifico dependen de las aproximaciones hechas durante la obtenci\'on de las EdE y, muy especialmente, de las ecuaciones de estructura.

Al utilizar las ecuaciones de estructura presentadas en el Cap\'itulo \ref{cap4}, se est\'a suponiendo que las Estrellas de condensado de Bose-Einstein son objetos est\'aticos esferiodales. Esto, en principio, permite determinar la masa, los radios polar y ecuatorial, el momento de inercia, el momento cuadrupolar y el corrimiento al rojo gravitacional.

\begin{figure}[!ht]
	\centering
	\includegraphics[width=0.49\linewidth]{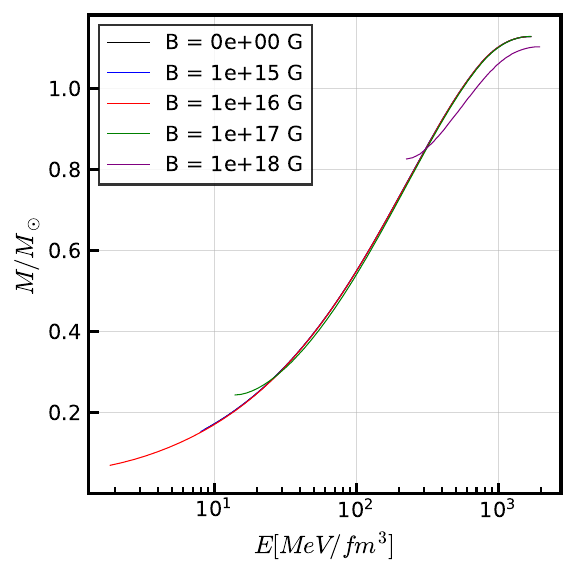}
	\includegraphics[width=0.49\linewidth]{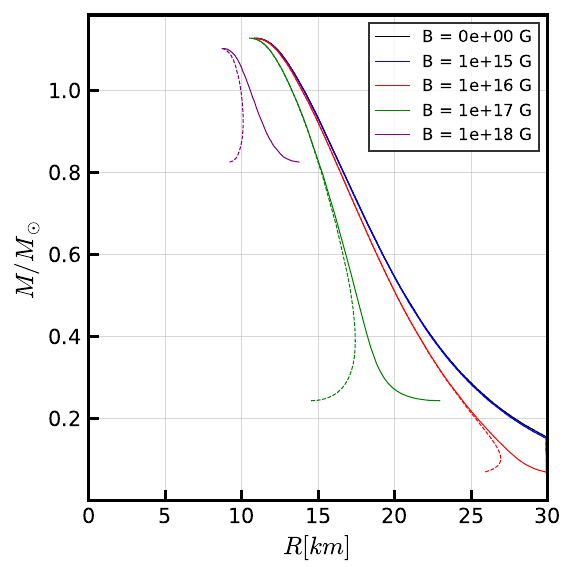}
	\caption{Los resultados de resolver las ecuaciones de estructura $\gamma$ para la EdE. Panel izquuierdo: la masa total de la estrella en funci\'on de la densidad de energ\'ia central. Panel derecho: relaciones masa-radio (lineas continuas: radio ecuatorial $r$, lineas discontinuas: radio polar $z$), para diferentes valores de $B$ y $U_0 = 10^9$ MeV/fm$^3$.}\label{mrzro}
\end{figure}

En la Fig.~\ref{mrzro} se muestran los resultados de integrar las ecuaciones de estructura $\gamma$ para $B=0$, $10^{15}$, $10^{16}$, $B=10^{17}$G y $B=10^{18}$G para las EdE~(\ref{edenpepi}) con la contribuci\'on de Maxwell. Como puede verse en el panel izquierdo, en el caso no magnetizado ($B=0$), la curva de $M$ vs $E$ muestra una regi\'on de configuraciones estelares estables cuya masa m\'axima es $M \approx 1.13 M_{\odot}$, con radio $R \approx 10.87$~km y densidad de energ\'ia central $\rho_0 \approx 169$~MeV/fm$^3$. Un campo magn\'etico constante cambia poco la forma de la curva $M$ vs $E$, pero disminuye ligeramente la masa m\'axima, aunque aumenta ligeramente la masa a menores densidades de energ\'ia.

La influencia del campo magn\'etico constante en la forma y tama\~no de las EBE es m\'as dram\'atica. El campo magn\'etico no solo deforma el objeto, sino adem\'as disminuye su tama\~no. Ambos efectos pueden verse en el panel derecho de la Fig.~\ref{mrzro}. Como $\gamma =z/r=P_{\parallel_0}/P_{\perp_0} $, el radio polar $Z$ es siempre menor que el ecuatorial $R$ porque $P_{\parallel}<P_{\perp}$. Esto significa que la estrella resultante es un objeto oblato. En la regi\'on de bajas densidades hay una enorme desviaci\'on con respecto a la curva de $B=0$. En esta region, la diferencia entre $R$ y $Z$ est\'a muy marcada, decreciendo esta con el aumento de la masa. Ambos radios decrecen con la densidad en lugar de tender a un valor constante. Este decrecimiento est\'a relacionado con la regi\'on de inestabilidad que muestran las EdE en forma tal que a medida que $P_{\perp_0}$ se acerca a cero, la estrella deviene menor. 

Las curvas mostradas en la Fig. (\ref{mrzro}) solo muestran las soluciones de las Ecs. (\ref{gTOVpartial}) consideradas estables de acuerdo a dos criterios: el primero est\'a relacionado con la estabilidad de la estrella respecto a oscilaciones radiales ($dM/dE>0$), lo que garantiza que el OC no se desintegre. El segundo criterio exige que la masa gravitacional de la estrella (calculada con las ecuaciones de estructura) sea menor que su masa bari\'onica, que es la suma de las masas de todas sus part\'iculas. Este \'ultimo criterio asegura la estabilidad del OC respecto a la dispersi\'on de la materia que lo compone; para aplicar este criterio partimos de la definici\'on de la masa bari\'onica $M_B$ \cite{Mariani2019hybrid}:
\begin{equation}
M_B = m_N \int_0^R \frac{4\pi r^2 n_B(r)}{[1 - 2Gm(r)/r]^{1/2}} dr,
\end{equation}

donde $m_N$ es la masa del neutr\'on y $n_B(r)$ es la densidad bari\'onica. Las regiones donde $M/M_B < 1$ son estables, mientras que aquellas donde $M/M_B > 1$ corresponden a configuraciones estelares inestables y deben descartarse. Sin embargo, todas las configuraciones estelares obtenidas fueron estables respecto a este criterio de estabilidad.

Los valores de masas m\'aximas obtenidos para dierentes valores del campo magn\'etico y par\'ametro de interacci\'on $U_0 = 10^9$ MeV.fm$^3$ se muestran en la Tabla \ref{MaxM}
\begin{table}[!ht]
\centering
\caption{Valores m\'aximos de las masas, con los radios correspondientes.}
\label{MaxM}
\renewcommand{\arraystretch}{1.3} 
\begin{tabular}{|c|c|c|c|}
\hline
\textbf{B (G)} & \textbf{$M$($M_{\odot}$)} & \textbf{R (km)} & \textbf{Z (km)} \\
\hline
$0$ & 1.12728 & 10.87000 & 10.87000 \\
\hline
$10^{15}$ & 1.12730 & 10.86000 & 10.85995 \\
\hline
$10^{16}$ & 1.12740 & 10.80947 & 10.80947 \\
\hline
$10^{17}$ & 1.12770 & 10.50348 & 10.50348 \\
\hline
$10^{18}$ & 1.10198 & 8.76026 & 8.76026 \\
\hline
\end{tabular}
\end{table}

\section{Estabilidad de las soluciones de las ecuaciones \texorpdfstring{$\gamma$}{gamma} y forma de las estrellas}

\begin{figure}[!ht]
	\centering
	\includegraphics[width=0.6\linewidth]{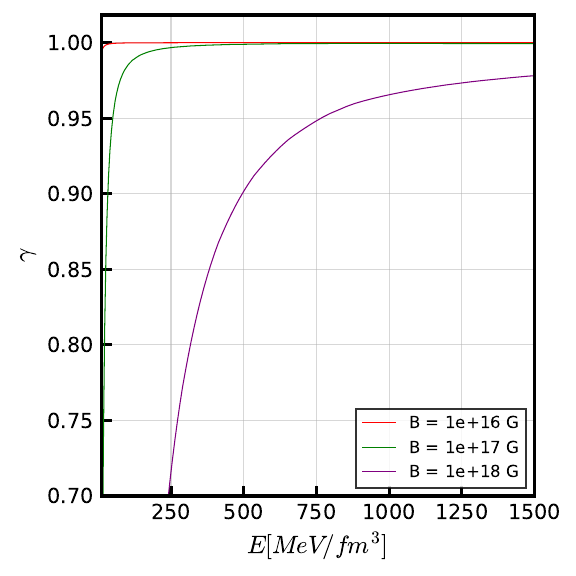}
	\caption{Comportamiento del par\'ametro $\gamma$ como funci\'on de la densidad de energ\'ia para diferentes alores de $B$.}\label{gammavse}
\end{figure}

La gran diferencia entre $R$ y $Z$ a bajas densidades se debe a que el par\'ametro $\gamma$ escapa del rango establecido\cite{WhiteGamma} ($1 > \gamma > 0.8$) para la validez de las Ecs. (\ref{gTOVpartial}). En la Figura (\ref{gammavse}) puede apreciarse el comportamiento del par\'ametro $\gamma$ para diferentes densidades de energ\'ia. Se aprecia que para mayores valores de $B$, el par\'ametro queda fuera del intervalo de estabilidad para rangos de $E$ cada vez mayores necesitando, para $B=10^{18}$ G energ\'ias mayores a $260$ MeV/fm$^3$ para entrar a la zona estable, lo que ya es muy superior a la correspondiente energ\'ia para $B=10^{17}$ G. Es interesante que otros modelos de EdE de Estrellas de Neutrones y ecuaciones de estructura tambi\'en presentan dificultades al tratar campos magn\'eticos del orden o superiores a $B = 10^{18}$ G. El límite te\'orico establecido por el teorema del virial para el campo magn\'etico m\'aximo que puede soportar una NS es precisamente $B \approx 10^{18}$ G \cite{Chandrasekhar1953}. Por lo tanto, nuestros resultados parecen apoyar este límite, indicando la relevancia de profundizar en la comprensi\'on de las razones físicas detr\'as del mismo.

Dado que los OCs esferoidales tienen dos radios principales, para la comparaci\'on definimos un radio medio $R_m$, de modo que la esfera que determina sea equivalente a la superficie de la estrella esferoidal:

\begin{equation}
A = 2\pi R \left[R + \frac{Z}{\epsilon} \arcsin \epsilon \right],
\end{equation}

donde $\epsilon = \sqrt{1+\gamma}$ es la excentricidad o elipticidad. De esta manera, el radio $R_m$ podría conectarse con la superficie de los objetos compactos y, en consecuencia, con sus propiedades de emisi\'on \cite{Samantha2024}. En el caso esf\'erico, $\gamma \rightarrow 1$ y $\epsilon \rightarrow 0$, mientras que para las estrellas m\'as deformadas $\gamma \rightarrow 0$ y $\epsilon \rightarrow 1$. En la Figura ~\ref{elipticity} mostramos la excentricidad como funci\'on de la densidad de energ\'ia central (panel izquierdo) y la masa (panel derecho).

\begin{figure}[!ht]
	\centering
	\includegraphics[width=0.49\linewidth]{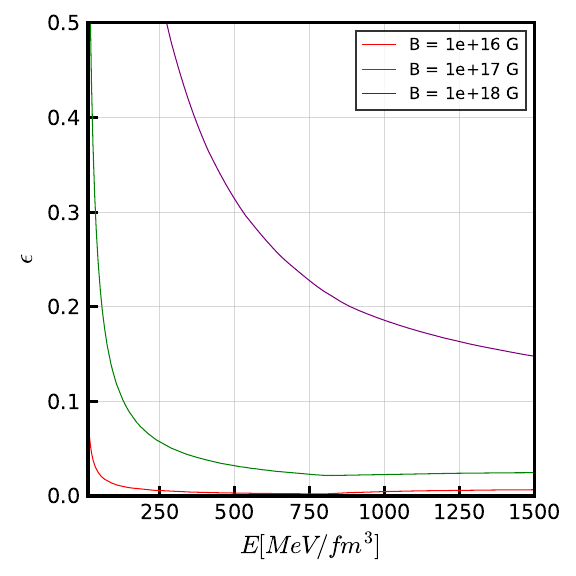}
	\includegraphics[width=0.49\linewidth]{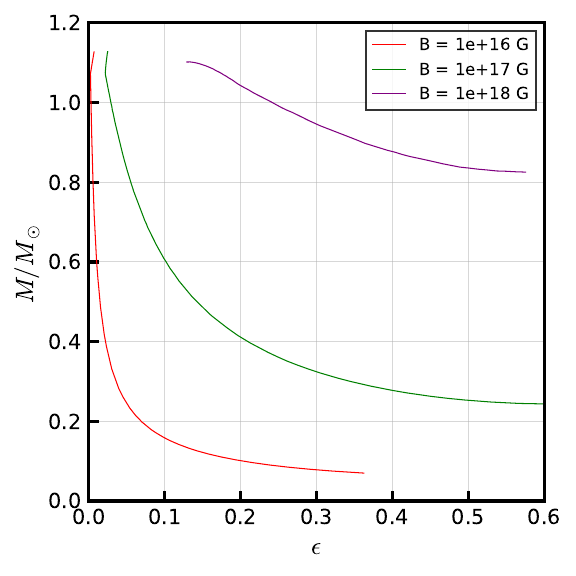}
	\caption{Excentricidad de la estrella como funci\'on de la densidad de energ\'ia central (panel izquierdo) y de la masa (panel derecho), para $B = 10^{16}, 10^{17}$ y $10^{18}$ G.}
\end{figure}\label{elipticity}

\section{Corrimiento al rojo gravitacional}

El corrimiento al rojo gravitacional es uno de los principales efectos predichos por la Teoría de la Relatividad General y, a su vez, constituye una de sus pruebas fundamentales. La frecuencia de un reloj at\'omico depende del valor del potencial gravitacional en el lugar de su ubicaci\'on. Por lo tanto, cuando un fot\'on es observado desde un punto con un potencial gravitacional m\'as alto, su longitud de onda se desplaza hacia el rojo.

Para un objeto compacto (CO) esferoidal est\'atico, el corrimiento al rojo est\'a dado por \cite{Terrero2021}:
\begin{equation}
z_{rs} = \frac{1}{\left(1 - \frac{2M}{R}\right)^{\gamma/2}} - 1,
\end{equation}
donde \( M \) y \( R \) son, respectivamente, la masa y el radio ecuatorial del CO. Si \( \gamma = 1 \), se recupera el caso esf\'erico.

\begin{figure}[!ht]
	\centering
	\includegraphics[width=0.6\linewidth]{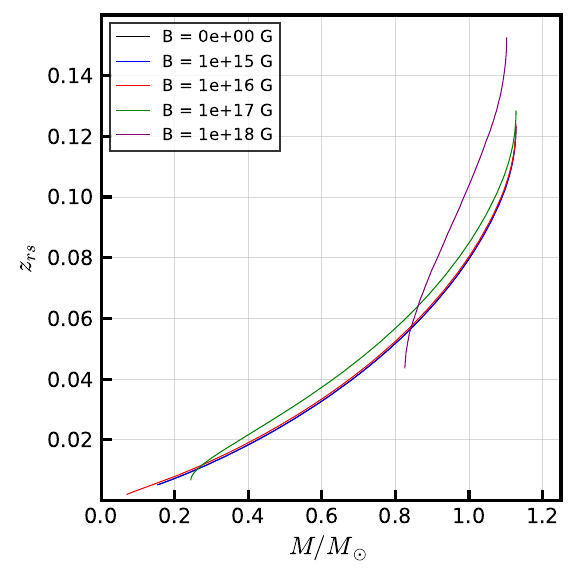}
	\caption{Corrimiento al rojo gravitacional como funci\'on de la masa de la estrella, para $B = 0, 10^{15}, 10^{16}, 10^{17}$ y $10^{18}$ G.}\label{Red}
\end{figure}

La definici\'on de \( z_{rs} \) incluye explícitamente el t\'ermino \( M/R \), por lo que su medici\'on puede restringir estos par\'ametros. Adem\'as, \( z_{rs} \) tiene un m\'aximo en el punto de masa m\'axima de la estrella, por lo que tambi\'en puede utilizarse para descartar EdE que no conduzcan a corrimientos al rojo observables y, por lo tanto, constituye un criterio para evaluar la viabilidad de los modelos de estrellas de neutrones (NSs) \cite{camenzind2007compact}. Nuestros resultados para \( z_{rs} \) se muestran en la Fig.~\ref{Red}.

Se aprecia que para campos magn\'eticos m\'as bajos (hasta $10^{16}$) apenas hay variaci\'on en $z_{rs}$ con respecto al caso no magnetizado. Para un campo $B = 10^{17}$ G, el valor de $z_{rs}$ comienza a ser apreciablemente mayor, mientras que para $B = 10^{18}$ G, el aumento del valor de $z_{rs}$ con la masa es notablemente mayor.

\section{Momento Cuadrupolar de Masa}
Finalmente, calculamos el momento cuadrupolar de masa de las estrellas esferoidales. La Relatividad General nos dice que esta magnitud est\'a directamente relacionada con la amplitud de las ondas gravitacionales (GWs) \cite{Samantha2024}, ya que estas solo se emiten en situaciones donde se genera una asimetría de masa que da lugar a un momento cuadrupolar. 

Matem\'aticamente, esto es consecuencia de emparejar las soluciones de las ecuaciones de campo de Einstein internas y externas, considerando condiciones de contorno en la superficie del objeto. Por lo tanto, la amplitud de la GW es proporcional a la variaci\'on en el tiempo del momento cuadrupolar de masa del objeto \cite{Samantha2024}. Así, se esperan GWs de eventos cataclísmicos como colisiones, rotaci\'on o campos magn\'eticos de objetos compactos que impliquen deformaciones lejos de la esfericidad. Por lo tanto, las estrellas esf\'ericas no tienen momento cuadrupolar y no pueden generar GWs. En contraste, las estrellas magnetizadas, descritas como esferoides, est\'an deformadas y tienen un momento cuadrupolar no nulo.

En el marco de nuestras ecuaciones de estructura, el momento cuadrupolar est\'a dado por\cite{ZubairiRomero_2015}:
\begin{equation}
Q = \frac{\gamma^3}{3} M^3 (1 - \gamma^2),
\end{equation}

donde \(\gamma = 1\) implica \(Q = 0\), correspondiendo al caso esf\'erico.
\begin{figure}[!ht]
	\centering
	\includegraphics[width=0.6\linewidth]{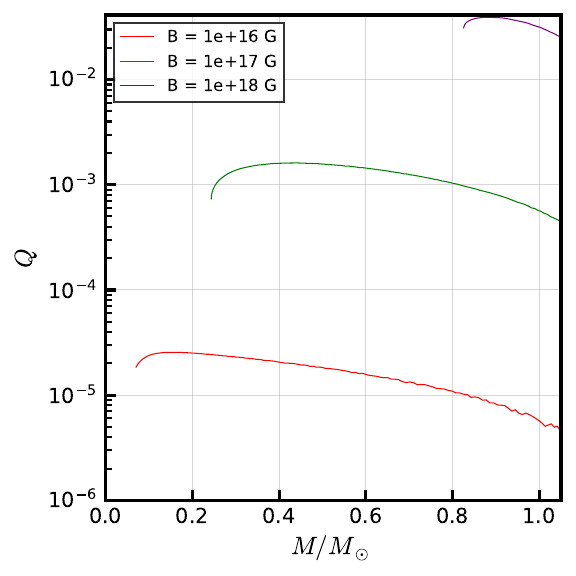}
	\caption{Momento cuadrupolar de masa vs masa total de la estrella, pata $B = 10^{16}, 10^{17}$ y $10^{18}$ G.}\label{Quad}
\end{figure}

La Fig.~\ref{Quad} muestra el momento cuadrupolar obtenido en funci\'on de la masa de la estrella. Las oscilaciones en la curva son un efecto de la presencia de la suma por los niveles de Landau en la EdE. $Q$ aumenta significativamente con $B$ y su m\'aximo se alcanza para estrellas en la regi\'on de masas y deformaciones intermedias. Este comportamiento se debe a la dependencia simult\'anea de \( Q \) en \( M \) y \( \gamma \) y, en particular, est\'a determinado por el hecho de que \( \gamma \) depende de la EdE y, por lo tanto, varía entre las estrellas. 

    \chapter*{Conclusiones}
\addcontentsline{toc}{chapter}{Conclusiones}
\label{Conclusiones}

La presente tesis ha estado dirigida a estudiar el efecto del campo magnetico en las EdE de estrellas de condensado de Bose-Einstein y en las ecuaciones de estructura de los mismos. En todos los casos hemos considerando un campo magnetico contante en la direccion z. Este trabajo constituye el primer estudio detallado de la contraparte bos\'onica cargada en condiciones extremas. A continuaci\'on, se resumen las principales conclusiones:

\begin{itemize}
    \item Se estudiaron las propiedades termodin\'amicas de un gas de bosones escalares cargados y fermiones en presencia de un campo magn\'etico uniforme, separando la contribuci\'on del vac\'io y la estad\'istica. En esta \'ultima, a su vez, se separ\'o la contribuci\'on del nivel de Landau m\'as bajo y de los niveles excitados. Para el estudio de estas propiedades termodin\'amicas, se prest\'o especial atenci\'on al límite de bajas temperaturas.
    \item Se estudi\'o el efecto del campo magn\'etico en las EdE de un gas mixto de electrones, protones, neutrones y piones cargados ($npe + \pi^{-}$) a temperatura cero, el cual induce anisotropías en las ecuaciones de estado, dividiendo la presi\'on en componentes paralela ($P_{\parallel}$) y perpendicular ($P_{\perp}$) al eje magn\'etico, efecto que es crucial para entender la estructura y estabilidad de los objetos compactos magnetizados.
    \item Se obtuvo un conjunto de condiciones de equilibrio para estrellas formadas por este gas (\textit{npe} + $\pi^-$), con las cuales se estudi\'o el comportamiento de los potenciales qu\'imicos de cada especie al variar la densidad bari\'onica del sistema para un campo magn\'etico fijo, as\'i como el valor de densidad para el cual comienzan a aparecer piones en el sistema para un rango de valores de la intensidad del campo magn\'etico.
    \item Se derivaron las ecuaciones de estado para un gas mixto de electrones, protones, neutrones y piones cargados ($npe + \pi^{-}$) a temperatura cero, mostrando que la presencia del campo magn\'etico modifica significativamente las propiedades del gas, especialmente en regímenes de baja densidad.
    \item Se resolvi\'o un sistema de ecuaciones de estructura para objetos compactos esferoidales, basado en una m\'etrica axisim\'etrica, la m\'etrica $\gamma$. Este formalismo permiti\'o modelar estrellas magnetizadas y calcular observables como la masa $M$, el radio $R$ y la deformaci\'on.
    \item Los resultados num\'ericos demostraron que el campo magn\'etico reduce la masa m\'axima de las EBE. Para campos intensos ($B \gtrsim 10^{18}$ G), se observ\'o inestabilidad, respaldando el límite te\'orico del teorema del virial.
    \item El corrimiento al rojo gravitacional $z_{rs}$ y el momento cuadrupolar $Q$ mostraron dependencia con la intensidad del campo magn\'etico, resultados que son relevantes para la interpretaci\'on de observaciones astron\'omicas.
\end{itemize}

Esta tesis ha proporcionado un marco te\'orico s\'olido para el estudio de estrellas de bosones magnetizadas, destacando el papel fundamental del campo magn\'etico en su estructura y estabilidad. Los resultados abren nuevas vías para la investigaci\'on en astrofísica de altas energías y física de partículas en condiciones extremas.

\chapter*{Recomendaciones}\addcontentsline{toc}{chapter}{Recomendaciones}
\label{Recomendaciones}

De acuerdo con los objetivos que nos trazamos al comenzar las investigaciones que culminan con la
presentaci\'on de esta tesis y con los resultados obtenidos en ella, formulamos, para la continuaci\'on del
trabajo, las siguientes recomendaciones

\begin{itemize}
    \item Extender el estudio a estrellas h\'ibridas con transici\'on de fase bos\'on - hadr\'on, obteniendo ecuaciones de estado h\'ibridas y cantidades asociadas.
    \item Incluir efector de temperaturas finitas para modelar etapas tempranas de evoluci\'on estelar.
    \item Continuar la b\'usqueda de m\'etricas y ecuaciones de estructura que permitan la descripci\'on de Objetos Compactos con anisotropía magn\'etica, sin la restricci\'on de que la deformaci\'on deba ser pequeña.
    \item Extender las ecuaciones de estructura al caso de Objetos Compactos anisotr\'opicos en rotaci\'on y comparar sus efectos con los del campo magn\'etico.
    \item Incorporar nuevas fases de la materia o interacciones m\'as complejas entre bosones y fermiones.
\end{itemize}

    \appendix
    \addcontentsline{toc}{chapter}{Apéndices}
    \chapter{Unidades y constantes físicas utilizadas}
\label{appA}

En la tesis se ha utilizado el sistema de unidades naturales (UN) para escribir todas la ecuaciones. En este sistema:

\begin{equation}\nonumber
\hbar=c=k_B = 1, 
\end{equation}

\begin{equation}\nonumber
[\text{longitud}]=[\text{tiempo}]= [\text{masa}]^{-1}=[\text{energía}]^{-1} =	[\text{temperatura}]^{-1},
\end{equation}
y
\begin{align} \nonumber
1 \text{ m} & = 5.07\times 10^{13} \text{ MeV}^{-1}, \\ \nonumber
1 \text{ kg} &= 5.61 \times 10^{29} \text{ MeV}, \\ \nonumber
1 \text{ s} &= 1.52 \times 10^{21} \text{ MeV}^{-1}, \\ \nonumber
1 \text{ K} &= 8.617\times10^{-11} \text{ MeV},
\\ \nonumber
1 \text{ J} &= 6.242\times10^{12} \text{ MeV},
\\ \nonumber
1 \text{T} &= 10^4 \text{ G} = 1.954 \times10^{-10} \text{MeV}^2.
\end{align}


En la Tabla \ref{tab:cons} se muestran las constantes físicas utilizadas en la tesis en unidades naturales. Adem\'as, usamos como referencia los principales par\'ametros del Sol:

\begin{table}[!ht]
	\begin{center}
		\begin{tabular}{ll@{}}
			\toprule
			Magnitud Física (Símbolo) 		          		&  UN  \\ \midrule
			Velocidad de la luz ($c$)    			    	& $1$			 \\
	        Constante de Dirac ($\hbar$) 		     		& 1  			\\
            Constante de Boltzmann ($k_B$)					& 1  			\\ \midrule	

            Masa del electr\'on ($m_e$) 				 	    & $0.511$ MeV  	 \\
            Masa del prot\'on ($m_p$) 				 	    & $938.272$ MeV  	 \\
            Masa del neutr\'on ($m_n$) 				 	    & $939.565$ MeV  	 \\			
            Masa del pion ($m_\pi$)
              & 139.570 MeV      \\
            Carga el\'ectrica del electr\'on ($e$)     	 		& $0.085$    	 \\
            Magnet\'on de Borh ($\mu_B$)                        & $0.083$ MeV$^{-1}$ \\
            Magnet\'on nuclear ($\mu_N$)                        & $4.528 \times 10^{-5}$ MeV$^{-1}$ \\ 
		  Momento magn\'etico an\'omalo del electr\'on ($\kappa$)
              & $8.662 \times 10^{-5}$ MeV$^{-1}$ \\
            Campo magn\'etico cr\'itico del electr\'on ($B_c^e$)
              & 3.070 MeV$^2$      \\
            Campo magn\'etico cr\'itico del prot\'on ($B_c^p$)
              & $10.357\times 10^{6}$ MeV$^2$      \\
            Campo magn\'etico cr\'itico del neutr\'on ($B_c^\pi$)
              & $229.174\times 10^3$ MeV$^2$      \\
            Campo magn\'etico cr\'itico del pion ($B_c^\pi$)
              & $229.174\times 10^3$ MeV$^2$       \\ \midrule	
              
			Constante de gravitaci\'on ($G$)                &  $6.711\!\times\!10^{-45}$ MeV$^{-2}$ \\
			Masa del Sol ($M_{\odot}$)    			     	 	    &  $1.116\!\times\!10^{60}$ MeV  	   \\
			Radio del Sol ($R_{\odot}$)     					    &  $3.528\!\times\!10^{21}$ MeV$^{-1}$
							
			\\ \bottomrule
		\end{tabular}
	\end{center}
	\caption{Constantes físicas utilizadas en la tesis expresadas en unidades naturales.}\label{tab:cons}
\end{table}. 

    \chapter{Integraci\'on de las ecuaciones de estructura anisotr\'opicas}
\label{appB}

Comenzando desde el centro de la estrella con densidad de energ\'ia $E_0 = E(r=0)$, y presiones centrales $P_{\parallel_0} = P_\parallel(r=0)$ y $P_{\perp_0} = P_\perp(r=0)$ tomadas de las ecuaciones de estado, las ecuaciones (\ref{gTOVpartial}) se hacen evolucionar hasta que una de las condiciones $P_\parallel (Z) = 0$ o $P_\perp (R) = 0$ se alcanza. Esto determina el radio correspondiente, ($R$ si $P_{\perp} = 0$ y  $Z$ si $P_{\parallel} = 0 $), a partir del cual el otro radio puede ser calculado a trav\'es de $Z = \gamma R$. Posteriomente, se eval\'ua la masa como $m=m(R)$. 

Para calcular la densidad de energ\'ia $E$ a partir de las EdE durante la integraci\'on de las Ecs.~(\ref{gTOVpartial}), denotemos por $c_1(N)$ y $c_2(N)$ a las curvas param\'etricas bidimensionales dadas por:
\begin{subequations}
	\begin{eqnarray}
	c_1(N)&=&(E(N),P_{\parallel}(N)) \label{c1}\\
	c_2(N)&=&(E(N),P_{\perp}(N)) \label{c2}
	\end{eqnarray}
\end{subequations}
con $E(N),P_{\parallel}(N)$ y $P_{\perp}(N)$ definidas por las EdE. Dados $\widetilde{P}_{\parallel}$ y $\widetilde{P}_{\perp}$, obtenidos en un paso de la integraci\'on de las Ecs.~(\ref{gTOVpartial}), dos valores param\'etricos, $\widetilde{N}_{\parallel}$ y $\widetilde{N}_{\perp}$ son calculados a partir de interpolar las EdE. Los puntos correspondientes en las curvas (\ref{c1}) y (\ref{c2}) son $c_1(\widetilde{N}_{\parallel})=(\widetilde{E}_{\parallel},\widetilde{P}_{\parallel})$ y $c_2(\widetilde{N}_{\perp})=(\widetilde{E}_{\perp},\widetilde{P}_{\perp})$, donde $\widetilde{E}_{\parallel}=E(\widetilde{N}_{\parallel})$ y $\widetilde{E}_{\perp}=E(\widetilde{N}_{\perp})$. Por tanto, en el pr\'oximo paso de la integraci\'on, el miembro derecho de la Ec.~(\ref{gTOVpartial2}) se actualiza utilizando el punto $c_1(\widetilde{N}_{\parallel})$ con $E=\widetilde{E}_{\parallel}$ y $P_{\parallel}=\widetilde{P}_{\parallel}$, mientras que el miembro derecho de Ec.~(\ref{gTOVpartial3}) se actualiza con $c_2(\widetilde{N}_{\perp})$ a partir de tomar  $E=\widetilde{E}_{\perp}$ y $P_{\perp}=\widetilde{P}_{\perp}$.

La existencia de dos valores de la densidad de energ\'ia en cada paso de integraci\'on, introduce la interrogante de cu\'al seleccionar a la hora de calcular la masa del objeto compacto Ec.~(\ref{gTOVpartial1}). Como estamos trabajando con un objeto anisotr\'opico, la variaci\'on de su densidad de masa tambi\'en debe ser diferente en las direcciones paralela y perpendicular al eje magnético. A lo largo de la direcci\'on ecuatorial, la densidad de masa es igual a:
\begin{eqnarray}\label{massdiff1}
dm &=& 4 \pi \gamma r^2 E_{\parallel} dr,
\end{eqnarray}
mientras que en la direcci\'on polar ella es:
\begin{eqnarray}\label{massdiff2}
dm &=& 4 \pi \frac{z^2}{\gamma^2} E_{\perp} dz.
\end{eqnarray}
En las Ecs.~(\ref{massdiff1}) y (\ref{massdiff2}) se han usado las densidades de energ\'ia paralela y perpendicular en dependencia de la direcci\'on en que se realiza la diferenciaci\'on. Si ahora tomamos en cuenta que $z = \gamma r$, la Ec.~(\ref{massdiff2}) puede transformarse en:
\begin{eqnarray}\label{massdiff3}
dM &=& 4 \pi \gamma r^2 E_{\perp} dr.
\end{eqnarray}

Sumando las Ecs.~(\ref{massdiff1}) y (\ref{massdiff3}), se obtiene:
\begin{equation}\label{massdifffinal}
\frac{dM}{dr} = 4 \pi \gamma r^2 \frac{E_{\parallel} + E_{\perp}}{2}.
\end{equation}

La Ec.~(\ref{massdifffinal}) indica que, si no se quiere perder informaci\'on acerca de la anisotrop\'ia en la densidad de masa, el lado derecho de la Ec.~(\ref{gTOVpartial1}) debe ser actualizado con la densidad de energ\'ia media $E=(\widetilde{E}_{\parallel}+\widetilde{E}_{\perp})/2$.

Finalmente, teniendo en cuenta lo que se acaba de discutir, el sistema de ecuaciones de estructura Ecs.~(\ref{gTOVpartial}) puede ser reescrito como:
\begin{subequations}\label{gTOV}
	\begin{eqnarray}
	&& \frac{dm}{dr}=4 \pi r^{2}\frac{(E_{\parallel} +E_{\perp})}{2}\gamma, \label{gTOV1}\\
	&&\frac{dP_{\parallel}}{dz}=\frac{1}{\gamma}\frac{dP_{\parallel}}{dr}=
	-\frac{(E_{\parallel}+P_{\parallel})[\frac{r}{2}+4 \pi  r^{3} G P_{\parallel}-\frac{r}{2}(1-\frac{2 G m}{r})^{\gamma}]}{\gamma r^{2}(1-\frac{2 G m}{r})^{\gamma}}, \label{gTOV3}\\
	&&\frac{dP_{\perp}}{dr}=-\frac{(E_{\perp}+P_{\perp})[\frac{r}{2}+ 4 \pi  r^{3} G P_{\perp}-\frac{r}{2}(1-\frac{2 G m}{r})^{\gamma}]}{ r^{2}(1-\frac{2 G m}{r})^{\gamma}}. \label{gTOV2}
	\end{eqnarray}
\end{subequations}

Las Ecs.~(\ref{gTOV}) contin\'uan acopladas entre s\'i a trav\'es de la masa y su dependencia en las densidades de energ\'ia paralela y perpendicular. N\'otese adem\'as que en el caso en que $\gamma=1$ las ecuaciones TOV son recuperadas.

En las Ecs.~(\ref{gTOV}), las coordenadas $z$ y $r$, est\'an ligadas a trav\'es del par\'ametro $\gamma$ que se ha supuesto constante, pero su valor espec\'ifico se desconoce, por lo que es necesario fijarlo de alguna manera. Desde el punto de vista matem\'atico y geom\'etrico, el par\'ametro $\gamma$ determina la deformaci\'on del objeto compacto, por tanto \'el debe estar relacionado de alguna manera con la anisotrop\'ia en las EdE, que es la causa f\'isica de la deformaci\'on. Como adem\'as $\gamma = z/r$, el problema ahora consiste en establecer una relaci\'on entre los radios polar y ecuatorial con las presiones en las direcciones paralela y perpendicular.

Teniendo en cuenta que en el caso isotr\'opico dada una EdE, las masas y radios est\'an un\'ivocamente determinados por los valores de densidad de energ\'ia y presi\'on en el centro de la estrella, es posible establecer la existencia de cierta proporcionalidad entre los radios y las presiones centrales correspondientes. La existencia de una dependencia entre el radio de una estrella y su presi\'on central es un hecho bastante bien establecido en la literatura, aunque su forma espec\'ifica, en general, se desconoce \cite{NSObserations}. Para enfrentar este problema, suponemos que la dependencia de los radios de la estrella con las presiones centrales es lineal, de manera que entonces $Z  \varpropto P_{\parallel0}$ y
$ R_{\perp}  \varpropto P_{\perp0}$. Teniendo en cuenta estas relaciones de proporcionalidad de conjunto con la Ec.~(\ref{gammazr}) se llega a:

\begin{equation}
\gamma=\frac{P_{\parallel_0}}{P_{\perp_0}}. \label{gammaa}
\end{equation}

La Ec.~(\ref{gammaa}) conecta la geometr\'ia del OC deformado con la anisotrop\'ia en las presiones del gas magnetizado que lo compone. Este ansatz implica que la estructura de la estrella est\'a solamente determinada por la anisotropía de la EdE en su centro e ignora la dependencia de la deformaci\'on con los perfiles internos de las presiones. Adem\'as, cuando $B = 0$, $P_{\parallel} = P_{\perp}$, $\gamma = 1$ y se recupera la soluci\'on isotr\'opica (ecuaciones TOV). 

En dependencia de la EdE, el par\'ametro $\gamma$ ser\'a: $ > 1$ ($P_{\parallel c} > P_{\perp c}$) para estrellas prolatas, $1$ ($P_{\parallel c} = P_{\perp c}$) para estrellas esféricas o $< 1$ ($P_{\parallel c} < P_{\perp c}$) para estrellas obl\'atas. El ansatz hereda la restricci\'on de que $\gamma$ debe ser cercano a 1 y esta limitaci\'on prueba ser esencial para obtener resultados razonables.

  \backmatter

  \bibliographystyle{ieeetr}  
  \bibliography{Thesis_Marcos} 	

\end{document}